\documentclass[12pt, letterpaper]{article}

\usepackage{amsmath}
\usepackage{amsfonts}
\usepackage{amssymb}
\usepackage{graphicx}

\pdfoutput=1

\usepackage{color}

\usepackage{amsmath, amssymb, graphics, epsfig, graphicx}
\usepackage{epsf}
\usepackage{epstopdf}
\usepackage {amssymb}
\newcommand{\nc}{\newcommand}
\nc{\ba}{\begin{eqnarray}}
\nc{\ea}{\end{eqnarray}}
\newcommand\be{\begin{equation}}
\newcommand\ee{\end{equation}}

\newcommand{\bea}{\begin{eqnarray}}
\newcommand{\eea}{\end{eqnarray}}

\def\bfk{{\bf k}}

\def\bfq{{\bf q}}

\def\Ylm{Y_{\ell q}(\theta, \phi) }
\def\ylm{Y_{\ell q}}

\def\zlm{Z_{\ell q}}

\def\Hm{H_{(m)} }

\newcommand{\bfx}{{\bf{x}}}
\newcommand{\bfm}{{{m}}}

\begin{document}

\vspace{5mm}
\vspace{0.5cm}
\begin{center}

\def\thefootnote{\fnsymbol{footnote}}

{ \bf  \large Cosmological Constant and Vacuum Zero Point Energy\\  in Black Hole Backgrounds }
\\[1cm]

{ 
Hassan Firouzjahi$\footnote{firouz@ipm.ir}$
}
\\[0.5cm]

{\small \textit{ School of Astronomy, Institute for Research in Fundamental Sciences (IPM) \\ P.~O.~Box 19395-5531, Tehran, Iran
}}\\

\end{center}

\vspace{.8cm}

\hrule \vspace{0.3cm}


\begin{abstract}

We study the quantum vacuum zero point energy in the Schwarzschild  black hole as well as in the Nariai limit of the dS-Schwarzschild backgrounds. 
We show that the regularized vacuum energy density near the black hole and also in the Nariai setup  match exactly with the corresponding value in the flat background, scaling with the fourth power of the mass of the quantum field.  The  horizon  radius of the dS space created from the vacuum zero point energy  introduces a new length scale which should be compared with the black hole horizon radius.
There is an upper limiting mass for the black hole immersed in the vacuum zero point energy which is determined by the mass of the Nariai metric associated to the dS background constructed from zero point energy. We calculate the variance in the distribution of the vacuum zero point energy and the density contrast  
and show that it develops strong  inhomogeneities on sub-horizon scales.


\end{abstract}
\vspace{0.5cm} \hrule
\def\thefootnote{\arabic{footnote}}
\setcounter{footnote}{0}
\newpage
\section{Introduction}

The cosmological constant problem is a  deep puzzle in theoretical cosmology. There have been numerous proposals for its resolution but none seems convincing \cite{Weinberg:1988cp, Sahni:1999gb, Peebles:2002gy, Copeland:2006wr, Martin:2012bt}.  On the other hand, the $\Lambda$CDM model has emerged as the Standard Model of Cosmology, very successful in describing the dynamics and the evolution history of the cosmos with about six free 
parameters  \cite{Weinberg:2008zzc, Planck:2018vyg}. 
Among unknown ingredients of the $\Lambda$CDM is the nature of dark energy which is taken to be just a constant term, the cosmological constant $\Lambda$, which seems to be the simplest and most economic choice. In this view, the cosmological constant problem has acquired a more direct relevance in physical cosmology.

To be more specific, there are actually two cosmological constant problems. The old cosmological constant problem is why the vacuum zero point energy density is not 
so large. Based on simple arguments, one may expect the scale of vacuum energy density $\rho_v$ to be at the order $M_{UV}^4$ in which $M_{UV}$ is a UV cutoff of high energy physics. If one takes $M_{UV}\sim M_P \sim 10^{18}$ GeV then there is a factor $10^{120}$ mismatch between the theoretical prediction and the observed value 
$\rho_\Lambda \sim 10^{-2}$ eV. The new cosmological constant problem is  why $\rho_\Lambda$ becomes comparable to the matter energy density at the current epoch in cosmic expansion history, at redshift $z\sim 0.3$.   

The cosmological constant problem was revisited recently in \cite{Firouzjahi:2022xxb} with the emphasis on the role of (A)dS horizon associated to the vacuum zero point energy.  It was argued that if the vacuum energy density is the sole distribution of energy then the spacetime filled with the vacuum zero point energy is unstable to quantum perturbations.   More specifically, the variance in the distribution of the vacuum energy density $\delta \rho$ was calculated and it was demonstrated  that the density contrast associated to the zero point energy is at the order unity, $\delta \rho/\langle \rho_v \rangle \sim 1$ indicating strong inhomogeneities in spacetime.  This conclusion was supported by the fact that the  dS horizon of the vacuum energy density associated to a heavy quantum field of mass $m$ is vastly smaller than the Hubble radius of the background FLRW universe. Therefore,  one can not expect to cover the entire cosmic background by too many uncorrelated tiny dS patches. It was argued that in order to prevent the strong inhomogeneities of zero point energy from destroying the background FLRW cosmology, one always requires a dominant classical source of energy density in the form of matter or radiation. As the universe expands and its horizon size increases, the dS patches of the heavy fields enter the horizon and generate strong inhomogeneities on sub-Hubble scales. 

In this work we extend the analysis of \cite{Firouzjahi:2022xxb} to the black hole backgrounds and calculate the vacuum zero point energy and its variance. This is a well-motivated question. Black  holes play important roles in the developments of theoretical physics and on the  understanding of quantum gravity. During the past decades  astronomers have suggested the existence of massive and supermassive black holes at the center of typical galaxies.  In addition, the recent detection of gravitational waves by the LIGO/Virgo/Advanced Virgo Collaboration team \cite{Abbott:2016blz, Abbott:2016nmj}  from the merging of binary astrophysical black holes have left no doubt on the reality of black holes in Nature.  So it is natural to study the vacuum zero point energy in a black hole background and examine if the conclusions reached in \cite{Firouzjahi:2022xxb} holds true in a black hole background as well.


\section{Preliminaries}
\label{prelim}

In this section we briefly review the standard analysis of calculating the vacuum zero point energy in connection to the cosmological constant problem which will be extended to the black hole backgrounds.

\subsection{The cosmological constant problem}

As in old cosmological constant problem, the trouble starts with the fact that the vacuum zero point energy can not be neglected in the presence of gravity. If one starts from a classical value of cosmological constant, then one can not ignore the contributions from the vacuum quantum  fluctuations. The latter's contributions are typically too large that either a severe fine-tuning or some fundamental symmetry principle are required to bring the effective values of the cosmological constant to its observed value.

To be specific, here we only consider a real scalar field with the mass $m$ but all physical conclusions can be extended to other fundamental fields such as fermions or 
vector bosons (gauge fields). To simplify the analysis we assume a free field with no interactions. The action of this quantum field in a curved background is given by 
\ba
\label{action-phi}
S=  \int d^4 x \, \sqrt{-g} \Big[ -\frac{1}{2} \partial^\mu  \Phi \partial_\mu \Phi
-  \frac{1}{2} m^2 \Phi^2 \Big] \, ,
\ea 
while the energy momentum tensor $T_{\mu\nu}$ has the following form,
\ba
\label{T-munu}
T_{\mu \nu} = \partial_\mu \Phi \partial_\nu \Phi - g_{\mu \nu} \Big( \frac{1}{2} \partial^\alpha \Phi \partial_\alpha \Phi + \frac{m^2}{2} \Phi^2
\Big)  \, .
\ea
Based on local Lorentz invariance and equivalence principle we expect that the vacuum expectation value $\langle T_{\mu \nu } \rangle$ to be proportional to the metric tensor,
\ba
\label{expectation}
\langle T_{\mu \nu } \rangle = -\langle \rho \rangle g_{\mu \nu} \, ,
\ea
in which $ \langle \rho \rangle$ is interpreted as the vacuum zero point energy density. Correspondingly, the vacuum pressure is expected to be 
$ \langle p \rangle = -  \langle \rho \rangle$. 

In conventional approach in dealing with the cosmological constant problem one imposes a hard momentum cutoff to regularize the UV divergence. As is well known in the literature, this violates the Lorentz invariance \cite{Martin:2012bt, Akhmedov:2002ts, Koksma:2011cq, Ossola:2003ku, Visser:2016mtr}. More specifically   the vacuum zero point energy density is given by
\ba
\label{rhov}
\langle \rho_v \rangle  = \frac{1}{2}\int \frac{ d^3 \bfk}{( 2 \pi)^3} \omega(k) 
  \,  ,
\ea
in which $\omega(k) \equiv \sqrt{\bfk^2 + m^2} $. The above integral is UV divergent so to read off the physical result it should be regularized. If one naively imposes the 
hard momentum cutoff $0\leq k \leq M_{UV}$, then one obtains the usually quoted results that $\langle  \rho_v \rangle \sim M_{UV}^4 $. However, the hard momentum cutoff  violates the underlying symmetry governing the system, i.e. the Lorentz invariance. Furthermore, it predicts a wrong equation of states so the expected results $\langle p_v \rangle = - \langle  \rho_v \rangle$ can not be 
obtained for the vacuum pressure.  One can bypass these difficulties by adding non-invariant counter terms in the corresponding regularization scheme. 
Alternatively,   one can employ a regularization scheme which is consistent with the underlying symmetry from the start.  
Using the standard dimensional regularization scheme, one actually obtains  \cite{Martin:2012bt, Akhmedov:2002ts, Koksma:2011cq}, 
\ba
\label{cc-formula1}
\langle \rho_v \rangle = -\langle p_v \rangle=   s\frac{\bfm^4}{64 \pi^2} \ln\big( \frac{\bfm^2}{\mu^2} \big)  \, , 
\ea
in which $\mu$ is the renormalization scale. Here  $s$ is the degree of polarization in which for a real scalar field  $s=1$, for a Diract fermion field $s=-4$ while for a massive vector field $s=3$. 

Some notable conclusions from the above formula is that the vacuum energy density scales with the fourth power of the mass of the quantum field. Second, massless fields such as gravitons, gluons and photons do not contribute to the vacuum energy density. Finally, depending on the energy scale of interest (i.e. 
$\mu)$ and the spin of the field, the vacuum energy density can be either positive (dS) or negative (AdS).  

The dS space associated with the vacuum energy density Eq. (\ref{cc-formula1}) has 
the expansion rate  $H_{(m)} \sim \frac{m^2}{M_P} $ and Hubble radius $H_{(m)}^{-1}$. For example, for the electron field with the mass $m_e \sim $ MeV the dS horizon is at the order $H_{(m_e)}^{-1} \sim 10^9 m$. If electron is the field responsible for the observed  cosmological constant at the present time with the Hubble radius $H_0^{-1}$, 
we need as many as  $\big( H_0^{-1}/H_{(m_e)}^{-1}\big)^3 \sim 10^{51}$ independent patches with the size $H_{(m_e)}^{-1}$ to cover the cosmos. This is unrealistic because the dS patches are expected to be uncorrelated as they are created quantum mechanically. Considering fields heavier than electron the situation worsen as the ratio $ H_0^{-1}/H_{(m)}^{-1}$ becomes much larger. Specifically, for a field of mass $m$ the number of independent patches to cover the background cosmos is 
\ba
\label{N-patch}
N_{\mathrm{patches}} \sim \Big( \frac{H_{(m)}}{H_0}\Big)^3 \sim \Big(\frac{m}{10^{-2} \mathrm{eV}} \Big)^6 \, .
\ea
Curiously we see for the neutrino with the mass at the order $10^{-2}$ eV we need only one patch to cover the entire cosmos. This suggests that the neutrino field maybe  behind the observed value of dark energy $\rho_\Lambda \sim (10^{-2})^3$ eV \cite{Firouzjahi:2022xxb}.   Note that neutrino is a fermion and its energy density has opposite sign compared to that of a boson. However, as discussed below Eq. (\ref{cc-formula1}), the sign of $\langle \rho_v \rangle$ depends on the ratio $m/\mu$ and the polarization factor $s$ so if we take $\mu > m_\nu$, then the contribution of the neutrino to the vacuum energy density is positive.

Of course the important question is if the heavy fields can not contribute to the observed dark energy then what roles in cosmology they will play? Obviously one can not ignore the large energy density $\sim m^4$ for a heavy field of mass $m$.
To answer this question, it was argued in \cite{Firouzjahi:2022xxb} 
that one should look at the variance  of the statistical distribution of the zero point energy. In other words, $\rho$ is a statistical 
field with an unknown statistical distribution with the mean value $\langle \rho_v \rangle$ given by Eq. (\ref{cc-formula1}). Like in all physical system, one needs information beyond the mean value of a statistical field to examine the distribution of the  field. More specifically, we need the variance $\delta \rho^2 \equiv \langle \rho^2 \rangle - \langle \rho \rangle^2$ to properly understand the physical property of the vacuum zero point energy. 

The variance of the vacuum zero point energy was calculated in \cite{Firouzjahi:2022xxb}. For the case of real scalar field it was demonstrated that 
\ba
\label{density-contrast}
\frac{\delta \rho_v}{\langle \rho_v \rangle} =\pm \sqrt{10} \, .
\ea
This suggests that the background constructed purely from the vacuum zero point energy is highly inhomogeneous in which the local regions inside each patch may collapse to black holes. We comment that the conclusion that the spacetime created from the vacuum zero point energy to be highly inhomogeneous was also reached in a series of papers 
\cite{Wang:2017oiy, Cree:2018mcx, Wang:2019mbh, Wang:2019mee}
where it was concluded that a uniform cosmological constant can not cover the large scale spacetime and  the local spacetime is very inhomogeneous as in Wheeler's spacetime foam.

However, now consider the total energy density $\rho_T \equiv \rho_v + \rho_F$ in which $\rho_F$ represents the classical energy density from the FLRW sources such as radiation and matter. Then performing the variance analysis we obtain 
\ba
\label{fraction} 
\frac{\delta \rho_T}{ \langle \rho_T \rangle} = \frac{\delta \rho_v}{\rho_F 
+ \langle \rho_v \rangle} =  \pm \sqrt{10}  \frac{  \langle \rho_v \rangle}{\rho_F 
+ \langle \rho_v \rangle} \, .
\ea
Now demanding that the total density contrast to be small, 
$| \frac{\delta \rho_T}{ \langle \rho_T \rangle} | <1 $,  we obtain 
$ \langle \rho_v \rangle  \lesssim \frac{ \rho_F}{2}$. This suggests that while the heavy fields have large energy density but at the same time they produce too much inhomogeneities on sub-Hubble scale. To be consistent with the requirement of having a stable cosmological background, the sub-Hubble inhomogeneities generated by heavy fields may collapse to black hole which may resolve the mystery of the origin of dark matter as well.

\subsection{The black hole background}
\label{black-hole-prelim}

Our goal in this work is to extend the above mentioned results  
to the case of the black hole background, see also   \cite{Birrell:1982ix, Parker:2009uva, Moreno-Pulido:2020anb, Moreno-Pulido:2021jmn, Moreno-Pulido:2022phq, Martin:2012bt} who studied the zero point energy in a curved background as well. It was shown in \cite{Martin:2012bt} that  Eq. (\ref{cc-formula1}), obtained in the flat background, does hold in a general curved background as well. This is physically expected since the vacuum energy density is a very local property. Based on equivalence principle, any curved background is locally like a flat spacetime so Eq. (\ref{cc-formula1}) is expected to hold in a curved background as well.  Having said this, however, it is a non-trivial exercise to verify this conclusion explicitly  in a black hole background. One important reason is that the black hole has an event horizon which separates the singularity from  an outside observer. Consequently, the notion of vacuum is a non-trivial question in this curved background. The vacuum defined by an observer in a nearly flat region far from the black hole is different than the vacuum for an observer near the event horizon. This is the main reason behind  the phenomena of Hawking radiation \cite{Hawking:1974sw}. Therefore, it worths to calculate the vacuum zero point energy in a black hole background and examines the physical consequences.
One important physical effect is that now we  have two competing horizon scales: the black hole event horizon and the horizon radius of (A)dS space associated with the vacuum energy density, $H_{(m)}^{-1}\sim M_P/m^2$. As we shall see, non-trivial effects emerge when these two horizon scales become comparable, parallel to the results obtained in \cite{Firouzjahi:2022xxb} in an FLRW background.

To simplify the analysis we consider the  Schwarzschild black hole with the metric 
\ba
\label{metric-Sch}
ds^2 = - (1- \frac{2GM}{r} ) dt^2 + \frac{dr^2}{(1- \frac{2GM}{r} )} + r^2 d \Omega^2,
\ea
in which $G$ is the Newton constant, 
$M$ is the mass of the black hole as measured by an observer at infinity,
$t$ is the time coordinate for the exterior region, $r$ is the radial coordinate and
$d \Omega^2 = d \theta^2 + \sin (\theta)^2 \, d \phi^2$ is the angular part of the metric represented by a two-sphere. To simplify the notation, we denote the angular coordinates collectively by $y^a$ for $a= \{\theta, \phi\}$ with the metric 
$\gamma_{ab} =   \mathrm{diag} \big(1, \sin(\theta)^2 \big)$.

Exploiting the rotational symmetry of the angular directions, we can expand the quantum field $\Phi$ in terms of the spherical harmonics  $Y_{\ell q} (\theta, \phi)$ as follows\footnote{We use the notation $Y_{\ell q}$  in place of the conventional 
notation $Y_{\ell m}$ of the spherical harmonics in order not to confuse the azimuthal label $m$ with the mass of the quantum field.} 
\ba
\label{Phi-Z}
\Phi = \frac{1}{r} \sum_{\ell =0}^\infty \sum_{ q=-\ell}^\ell   Z_{\ell q} Y_{\ell q} (\theta, \phi)  \, ,
\ea
in which $Z_{\ell q}$ will play the role of the quantum mode operator. 
Also note that we have pulled out a factor $1/r$ such that $Z_{\ell q}$ will be
the canonically normalized field as we shall see below. 
The spherical harmonics satisfy the following relations which will be used frequently in our following analysis:
\ba
\label{sum-Ylm}
Y_{l q}(\theta, \phi)^* = (-1)^q  Y_{l \,  -q}(\theta, \phi) \, ,
\quad \quad 
\sum_{q=-\ell}^{\ell} | Y_{\ell q} (\theta,\phi)|^2 =\frac{2 \ell +1}{4 \pi} \, ,
\ea
with the normalization condition
\ba
\int d \Omega\,  \Ylm Y_{\ell' q'}^*(\theta, \phi) = \delta_{\ell \ell'} \delta_{q q'} \, . 
\ea
The reality condition of $\Phi$, $\Phi= \Phi^*$, requires that $Z_{\ell q}^* = (-1)^q Z_{\ell  \, -q}$. 

Now defining  the tortoise coordinate $d r_* = \frac{dr}{(1- \frac{2G M}{r}) }$, the action of the scalar field takes the following canonical form
\ba
\label{action-canonic}
S=\frac{1}{2} \int dt dr_* \sum_{\ell, q}
\Big[  \big | \frac{\partial Z_{\ell q}}{\partial t} \big|^2
- \big | \frac{\partial Z_{\ell q}}{\partial r_*} \big|^2 - 
\big(1- \frac{2G M}{r} \big) \Big( \bfm^2 + \frac{\ell (\ell +1) }{r^2} + \frac{2 G M}{r^3}\Big)
|Z_{\ell q}|^2
\Big] \, .
\ea
From the above action one obtains the standard Regee-Wheeler perturbation equation
\ba
\label{eq-tort-1}
\partial_{r_*}^2  \zlm - \partial_t^2 \zlm  -   \big(1- \frac{2 G M}{r} \big)  \left[\bfm^2  + \frac{\ell (\ell+1)}{r^2} + \frac{2 G M}{r^3}
\right]  \zlm =0  \, .
\ea
As the time coordinate enjoys the translation invariance we can take 
$\zlm \propto e^{-i \omega t}$ yielding the following Schrodinger-like equation
\ba
\label{eq-tort}
\partial_{r_*}^2  \zlm + \big( \omega^2 - V_{\mathrm{eff}} \big) \zlm = 0 \, ,
\ea 
with the effective potential
\ba
\label{Veff}
V_{\mathrm{eff}} = \big(1- \frac{2 G M}{r} \big)  \left[\bfm^2  + \frac{\ell (\ell+1)}{r^2} + \frac{2 G M}{r^3} \right] \, .
\ea 
Unfortunately the above equation can not be solved analytically to obtain the mode function. Therefore, we solve for the mode function in two extreme regimes, near the horizon region $r \simeq 2 G M$ and for the region far from the black hole, $r \gg 2 GM$.

\section{Zero Point Energy Near Horizon}
\label{near-horizon}

Here we calculate the vacuum zero point energy near the horizon region 
$r \simeq  r_S $ with $ r_S \equiv 2 GM$ representing the Schwarzschild radius of the  black hole. 

As is well known the $(t, r)$ coordinate fails to cover the entire black hole manifold. Specifically, near the horizon the $(t, r)$ coordinate is singular and one can not use it to perform physically meaningful analysis. Instead, we can use the Kruskal coordinate $(T, R)$ which covers the entire manifold and is regular on the surface of event horizon.  

Going to Kruskal coordinate, the metric (\ref{metric-Sch}) takes the following form 
\ba
\label{metric-Kruskal}
ds^2 =  \frac{32 G^3M^3}{r} e^{-r/2GM} ( -d T^2 + d R^2) +   r^2 d \Omega^2
\ea
in which $(T, R)$ coordinate is related to the original  coordinate $(t, r)$ via
\ba
\label{UV-r}
T^2 - R^2 =  e^{r/2GM} ( 1- \frac{r}{2GM} )  \, ,
\ea
and
\ba
\label{UV-t}
\frac{T}{R}  = \tanh(\frac{t}{4 G M}) \, .
\ea
The good thing about the $(T, R)$ coordinate is that it is non singular near the horizon while the metric being conformally flat. 

Near horizon the metric (\ref{metric-Kruskal}) takes the locally flat form
\ba
\label{metric-near}
ds^2 (r \rightarrow r_S) = - d \tau^2 + dx^2 + r_S^2 d \Omega^2 \, .
\ea
in which $(\tau, x)$ are the local cartesian coordinates which are simply related to
the $(T, R)$ coordinate near $r=r_S$ by the simple rescaling 
$(\tau, x) = 4 G M e^{-1/2}   (T, R) $.  Now the metric (\ref{metric-near}) is flat with the structure of $R^2 \times S^2$ in which the angular parts of the manifold are restricted to a two-sphere of radius $r_S$.  Since the spacetime near horizon is locally  flat then one logically expects that the vacuum zero point energy to agree with its flat value.  Here we demonstrate it explicitly. However, the crucial point is that the vacuum defined for the locally inertial  observer near the horizon is the vacuum associated to the Kruskal coordinate denoted by $|0\rangle_K$. This vacuum is different than 
the vacuum employed by an observer for regions far from black hole. The difference in the notion of vacuum as measured by these two observers is the main reason behind the Hawking radiation \cite{Hawking:1974sw}. Correspondingly the physical vacuum zero point energy near the horizon is the one measured by the Kruskal observer (i.e. the locally inertial observer).  To prevent confusions, we denote the expectation value of the zero point energy as measured by the Kruskal observer by 
${\langle \rho_v \rangle}_K$. 

Using the decomposition (\ref{Phi-Z}) the action in the $(\tau, x)$ coordinate takes the following simple form
\ba
\label{action-near} S= \frac{1}{2} \int d \tau dx \sum_{\ell q} \Big[
\big | \frac{\partial Z_{\ell q}}{\partial \tau} \big|^2 -  \big | \frac{\partial Z_{\ell q}}{\partial x} \big|^2 - {\bf m_\ell}^2 |\zlm|^2 \, 
\Big] \, ,
\ea 
in which ${\bf m_\ell}$ is the effective mass given by ${\bf m_\ell}^2 \equiv \bfm^2 + \frac{\ell (\ell +1)}{r_S^2}$ which appears because the angular parts of the manifold are confined on a two-sphere of radius $r_S$.  Correspondingly, the mode function satisfies the following simple equation
\ba
\partial_\tau^2 \zlm - \partial_x^2 \zlm +{\bf m_\ell}^2 \zlm =0 \, .
\ea
Now expanding $\zlm \propto e^{- i \omega_\ell \tau+ i k x} $ we have the relation
$\omega_\ell = \sqrt{k^2 + {\bf m_\ell}^2}$ for the positive frequency mode. 

To quantize the system, we expand the mode function in terms of the annihilation and  creation operators $a_{\bfk}^{\ell m}$ and ${a_{\bfk}^{\ell m}}^\dagger$ which satisfies the following commutation relations
\ba
\label{commute}
\big [  a_{\bfk}^{\ell q} , {a_{\bfk'}^{\ell' q'}}^\dagger \big]  = \delta_{\ell \ell'} \delta_{q q'}
\delta(\bfk- \bfk') \, .
\ea
Note that since we have a two-dimensional quantum field (as represented by the action (\ref{action-near}))  spanned by the coordinate $(\tau, x)$, then the momentum $\bfk$ is one-dimensional representing the Fourier expansion for the $x$-coordinate.    Imposing the quantum commutation relation between the field $\Phi$ and its conjugate momentum $\partial_\tau \Phi$ we obtain  the quantum mode function as follows
\ba
\label{Z-mode-near}
\Phi = \frac{1}{r_S} \int \frac{d \bfk}{\sqrt{2 \pi}} \sum_{\ell q}
\frac{1}{\sqrt{2\omega(k)}}  
\Big[ e^{i k\cdot x} a_\bfk^{\ell q} + (-1)^q e^{-i k\cdot x} {a_\bfk^{\ell\,  -q}}^\dagger
 \Big] \ylm \, .
\ea 
Here $k^\mu = (\omega_\ell, \bfk)$ with $x^\mu = (\tau, x)$. Note that the factor $(-1)^q$ in the above expansion appeared from the reality condition
that $\zlm^* = (-1)^q Z_{\ell \, -q}$. 

The vacuum energy density $\rho_v$ is given by  
\ba
\rho_v= \frac{1}{2} m^2 \Phi^2 + \frac{1}{2} (\partial_\tau \Phi)^2 +   \frac{1}{2} (\partial_x \Phi)^2
+ \frac{1}{2 r_S^2} \gamma^{a b} \partial_a \Phi \partial_b \Phi \, ,
\ea
in which $\gamma_{ab}$ represents the metric on a unit two-sphere.  Denoting the above four components respectively by $\rho_i$, for i=1...4, we then have $\langle \rho \rangle_K = \sum_{i}^{4} \langle \rho_i \rangle_K$.  Now we calculate  $\langle \rho_i \rangle$ in turn. 

We start with $\langle \rho_1 \rangle_K = \frac{m^2}{2} \langle \Phi^2 \rangle_K $ which is the easiest. Using the mode function (\ref{Z-mode-near}) and the commutation relation 
Eq. (\ref{commute}) and the summation relation (\ref{sum-Ylm})
we obtain 
\ba
\label{rho1K}
\langle \rho_1 \rangle_K=  \frac{m^2}{2} \langle \Phi^2 \rangle_K =
\frac{m^2}{16 \pi r_S^2} \sum_{\ell=0}^{\infty} ( 2 \ell +1) \int_{-\infty}^{\infty} \frac{d \bfk}{2 \pi} \frac{1}{\omega_\ell(k) } \, ,
\ea
in which $\omega_\ell^2 = k^2 + \bfm^2+  \frac{\ell (\ell +1)}{r_S^2}$.   

Compared  to  the flat four-dimensional  case, there are a few notable differences for the integral of the zero point energy above. First, the integral over the Fourier mode $\bfk$ is one-dimensional. Second, the effective mass depends on the quantum number $\ell$
and we have to perform an infinite sum over $\ell$. The regularization of the above integral, specially the summation over $\ell$, is somewhat non-trivial which we elaborate in some details below. 
  
 To perform the integral over $\bfk$ we employ the dimensional regularization scheme  by replacing $d \bfk/(2 \pi)  \rightarrow  \mu^{2-d} k^{d-2} dk/(2 \pi)^{d-1}$
 in which $\mu$ is a mass scale to keep track of the mass dimension of the energy density. After performing the integral over $k$ for arbitrary value of $d$, we consider 
 $d= 2- \epsilon$ for small value of $\epsilon $ to regularize the singular $1/\epsilon$ term and read off the physical finite terms.  
 
 Performing the dimensional regularization for the integral over $k$, we have
 \ba
 \label{rho1-1}
 \langle \rho_1 \rangle_K &=& 
 \frac{2  \mu^{2-d} m^2}{16 \pi r_S^2} \sum_{\ell=0}^{\infty} ( 2 \ell +1) \int_{0}^{\infty} 
 \frac{d k}{(2 \pi)^{d-1}}  \frac{k^{d-2}}{ \sqrt{k^2 + {\bf m_\ell}^2} } \nonumber\\
 &=&
 \frac{m^2 2^{1-d} \pi^{-d+\frac{1}{2}} \mu^{2-d}}{16 \pi r_S^d} \Gamma(1- \frac{d}{2} ) 
 \Gamma(\frac{d-1}{2}) \,  { S}_1 \, ,
 \ea
 in which  ${ S}_1$ is the following  infinite sum 
 \ba
 \label{S1-def}
 { S}_1 \equiv \sum_{\ell=0}^{\infty} ( 2 \ell + 1) \Big(  \kappa^2 + \ell (\ell+1)
 \Big)^{\frac{d-2}{2}} \, ,
 \ea
 where $\kappa \equiv m r_S$.  
 
 We are interested in the physical limit where the Compton length of the fundamental particle is much smaller than the black hole 
 horizon, $m^{-1} \ll r_S$, so in this limit of interest $\kappa \gg1$.  Expressed in terms of the mass of black hole, we assume a massive enough black hole, satisfying the condition $m M \gg M_P^2$. 
  
It is challenging to calculate the above infinite sum. Happily, we were able to find an analytic expression for it (for more details see Appendix \ref{Sum-reg}), yielding
 \ba
 \label{S1-val}
  {\Large S}_1 =  -\frac{2}{d} \kappa^d \, .
 \ea 
 It is remarkable that the infinite sum (\ref{S1-def}) can be expressed in the above simple form.
 
 Plugging  the above value of $ {\Large S}_1$ into the expression Eq. (\ref{rho1-1}) and then performing the expansion $d= 2-\epsilon$ to leading order in $\epsilon$ and regularizing the $\frac{1}{\epsilon}$ singular term, we obtain the following finite value
 for $ \langle \rho_1 \rangle_K$:
 \ba
 \label{rho1}
  \langle \rho_1 \rangle_K = \frac{\bfm^4}{32 \pi^2} \ln \Big( \frac{\bfm^2}{\mu^2} \Big) \, ,
 \ea
in which now $\mu$ is the renormalization scale.   We note that in obtaining the above finite value, we have performed actually two back to back regularizations.
First, we have to regularize the infinite sum in  $ { S}_1$ to read off its
finite value. This was done by analytic continuation as the sum defined in
Eq. (\ref{Sp}) is convergent only for $p>1$ which for our case $ { S}_1$ 
in Eq. (\ref{S1-def}) corresponds to $d<0$. Second was the conventional dimensional regularization to regularize the divergent $\frac{1}{\epsilon}$ contribution from  the final answer.

Performing the same steps for $\langle \rho_2 \rangle_K$ we obtain
\ba
\label{rho2-1}
\langle \rho_2 \rangle_K = \frac{1}{2}\langle (\partial_\tau \Phi)^2 \rangle_K
&=&
\frac{1}{16 \pi r_S^2} \sum_{\ell=0}^{\infty} ( 2 \ell +1) \int_{-\infty}^{\infty} \frac{d \bfk}{2 \pi} {\omega_\ell(k) } \,, \\
&=&
-\frac{ 2^{-d} \pi^{-d+\frac{1}{2}} \mu^{2-d}}{16 \pi r_S^{d+2}} \Gamma(- \frac{d}{2} ) 
 \Gamma(\frac{d-1}{2})\,  { S}_2 \,  \nonumber,
\ea
in which
 \ba
 { S}_2 \equiv \sum_{\ell=0}^{\infty} ( 2 \ell + 1) \Big(  \kappa^2 + \ell (\ell+1)
 \Big)^{\frac{d}{2}} = -\frac{2 \kappa^{d+2}}{d+2} \, .
 \ea
Plugging the above value of ${ S}_2$ in Eq. (\ref{rho2-1}) and after regularization, we obtain
\ba
 \label{rho2}
  \langle \rho_2 \rangle = \frac{\bfm^4}{128 \pi^2} \ln \Big( \frac{\bfm^2}{\mu^2} \Big) \, .
 \ea

Similarly, for $\langle \rho_3 \rangle_K$ we obtain
\ba
\label{rho3-1}
\langle \rho_3 \rangle_K = \frac{1}{2}\langle (\partial_x \Phi)^2 \rangle_K
&=&
\frac{1}{16 \pi r_S^2} \sum_{\ell=0}^{\infty} ( 2 \ell +1) \int_{-\infty}^{\infty} \frac{d \bfk}{2 \pi} \frac{k^2}{\omega_\ell(k) } \, ,\\
&=&
\frac{ 2^{1-d} \pi^{-d+\frac{1}{2}} \mu^{2-d}}{16 \pi r_S^{d+2}} \Gamma( \frac{1+d}{2} ) 
 \Gamma(-\frac{d}{2})\,  { S}_2 \, \nonumber,
\ea
yielding to the following finite value
\ba
 \label{rho3}
  \langle \rho_3 \rangle_K = -\frac{\bfm^4}{128 \pi^2} \ln \Big( \frac{\bfm^2}{\mu^2} \Big) \, .
 \ea

The computation of $\langle \rho_4 \rangle_K$ is somewhat non-trivial as we have the additional contribution $\ell (\ell+1)$ in the infinite sum. More specifically, we have
\ba
\label{rho4-1}
\langle \rho_4 \rangle_K = \frac{1}{2 r_S^2}\gamma^{a b}\langle \partial_a \Phi  \partial_b \Phi \rangle_K
&=&
\frac{1}{16 \pi r_S^4} \sum_{\ell=0}^{\infty} ( 2 \ell +1)  \ell (\ell+1) \int_{-\infty}^{\infty} \frac{d \bfk}{2 \pi} \frac{1}{\omega_\ell(k) } \,,  \\
&=&
\frac{ 2^{1-d} \pi^{-d+\frac{1}{2}} \mu^{2-d}}{16 \pi r_S^{d+2}} \Gamma( \frac{d-1}{2} ) 
 \Gamma(1-\frac{d}{2})\,  { S}_4 \,  \nonumber,
\ea
in which (see Appendix \ref{Sum-reg} for further details)
\ba
\label{hatS}
{S}_4 \equiv \sum_{\ell=0}^{\infty} ( 2 \ell + 1) \ell (\ell+1)  \Big(  \kappa^2 + \ell (\ell+1) \Big)^{\frac{d-2}{2}}  = \frac{4 \kappa^{d+2}}{d(d+2)} \, .
\ea
Plugging the above value of ${ S_4}$ into Eq. (\ref{rho4-1}) and performing the regularization yields
\ba
\label{rho4}
  \langle \rho_4 \rangle_K = -\frac{\bfm^4}{64 \pi^2} \ln \Big( \frac{\bfm^2}{\mu^2} \Big) \, .
\ea

Summing all contributions, we finally obtain
\ba
\label{rho-val} 
\langle \rho \rangle_K= \sum_i^4 \langle \rho_i \rangle = \frac{\bfm^4}{64 \pi^2} 
\ln \Big( \frac{\bfm^2}{\mu^2} \Big) \, .
\ea
Interestingly, the above result matches exactly with the value obtained in the flat background.  Also we see that
the dependence on the horizon radius $r_S$ has dropped from the final answer. 
This is consistent with the expectation that the vacuum energy density is a very local property so it can not depend on large scale structure of the manifold such as 
the horizon radius. However, to reach this conclusion, it was essential to use the Kruskal vacuum $|0\rangle_K$ as discussed before. 

Also looking at various contributions  $\langle \rho_i \rangle$ we have the following relations among them which will be useful in our upcoming analysis for the variance and the density contrast, 
\ba
\label{rho-i-relation0}
\langle \rho_1 \rangle = 4 \langle \rho_2 \rangle =-4 \langle \rho_3 \rangle =
-2 \langle \rho_4 \rangle = 2 \langle \rho \rangle \, .
\ea

In the above analysis we have calculated $\langle T_{00}\rangle_K = \langle \rho \rangle_K$. Calculating  the expectation of other components  
$\langle T_{\mu \nu}\rangle_K$ one can show that indeed the relation  (\ref{expectation}) does hold. 

We comment that our analysis of calculating the quantum expectation of the vacuum zero point energy $\langle \rho_v \rangle$ has some relevance 
to the idea of ``vacuum polarization" by gravitational effects in black hole backgrounds  which is studied extensively in the past, see for example \cite{Candelas:1980zt, Howard:1984qp, Frolov:1982fr, Frolov:1983ig, Anderson:1990jh, Anderson:1994hg}. In these analysis the renormalized value of $\langle \Phi^2 \rangle$ and $\langle T^\mu_\nu \rangle$ in a black hole background  were calculated in which, to get rid of the QFT infinities, the point splitting regularization scheme is usually employed \cite{DeWitt:1975ys, Fulling:1989nb, Christensen:1976vb,Christensen:1978yd}. For example,  $\langle T^0_0 \rangle$  for a scalar field of large mass with $\kappa =m r_S\gg 1$ in  the Hartle-Hawking vacuum is obtained 
to be  \cite{Frolov:1982fr} $\langle T^0_0 (r)\rangle= ( c_1+ c_2 \frac{M}{r}) \frac{G^2 M^2}{m^2 r^8}$ with some known numerical factors $c_1$ and $c_2$. In particular, near the horizon $r= r_S=2 GM$, one obtains 
$\langle T^0_0 \rangle \sim \frac{M_P^{12}}{m^2 M^6} \propto \frac{1}{m^2}$. 
In performing the point splitting regularization, all divergent terms containing non-negative powers of $m$, including terms containing $m^4, m^2, m^0$ and $\ln(m)$ have been removed. In addition, the leading contribution to cosmological constant term  scaling like $m^4 \ln(m)$  is discarded. In comparison, in our analysis we have kept only the leading order in $\kappa$ in the series expansion such as in ${ S_1}$. However, we have checked that if we include the subleading terms of oder $\kappa^{-2}$ in 
the series expansion such as in ${ S_1}$ we can also generate the results starting at the order $\frac{1}{m^2}$ in $\langle \Phi^2 \rangle$ and  
$\langle T^0_0 \rangle$  as in \cite{Frolov:1982fr, Anderson:1990jh}. In particular, 
calculating the required subleading terms in $S_1$ (see Appendix \ref{Sum-reg} for further details) we have obtained the first 
finite term in $\langle \Phi^2 \rangle$  at $r= r_S$ to be   $\langle \Phi^2 \rangle= \frac{1}{240 \pi^2} \frac{1}{m^2 r_S^4}$ in agreement with the previous literature. 

Before closing this subsection, we comment about the dimensional  regularization scheme used here. As emphasized before, dimensional regularization scheme has the advantage that it respects the underlying local Lorentz invariance so the 
regularized physical results have the covariant form. However, one may use other scheme to perform the regularization such as the well established zeta function regularization. In these schemes one encounters power law divergences, such as quartic and quadratic divergences,   which should be removed via analytic continuation. While the quartic  and quadratic divergences respectively have the equation of state like radiation and spatial curvature, only the logarithmic terms have the proper equation of state of the vacuum \cite{Martin:2012bt}. Also note that here we deal with UV divergences in which it is known that the dimensional regularization is applicable. However, care must be taken in the case of infrared divergences in which additional physical inputs have to be added, 
see for example \cite{Weinberg:1995mt}.

\subsection{Variance of zero point energy density}
\label{variance-near}

In the above analysis the average vacuum energy density $\langle \rho_v \rangle_K$ has been calculated. However, as advocated in 
\cite{Firouzjahi:2022xxb},  this does not mean that the vacuum energy is statistically uniformly distribute. Indeed one should think of the field perturbations $\Phi$ and the energy density $\rho$ as random fields which can have non-trivial statistical distributions. So calculating the mean value $\langle \rho \rangle$ is not enough to conclude that we have a uniform space-filling energy distribution. We also have to calculate the variance of the vacuum 
energy distribution  $\delta \rho^2 \equiv \langle \rho^2 \rangle - \langle \rho \rangle^2$ and compare its value to the mean value. If we encounter a situation where the density contrast $\delta \rho/\langle \rho \rangle \sim 1$, then the background covered by the vacuum zero point energy  is actually very inhomogeneous.

Here, following  \cite{Firouzjahi:2022xxb},  we calculate the variance of vacuum zero point energy for near horizon regions using the Kruskal vacuum.  More specifically,
\ba
\delta \rho^2_K = \langle \rho^2 \rangle_K - \langle \rho \rangle_K^2 = 
\sum_i^4 \big(  \langle \rho_i^2 \rangle_K  - \langle \rho_i \rangle_K^2 \big) \, .
\ea
In obtaining the last result, we have used the relation 
$\langle \rho_i \rho_j \rangle_K = \langle \rho_i \rangle_K \langle \rho_j \rangle_K$
which can be verified using the form of the mode function given in 
Eq. (\ref{Z-mode-near}). 

Using the Gaussian structure of $\Phi$ one can show that
\ba
\label{Wick1}
\langle \rho_i^2 \rangle_K = 3 \langle \rho_i \rangle_K^2 \, , 
\quad \quad i=1,2,3 ,
\ea
which is a realization of the Wick contractions. We demonstrate the above relation more explicitly in Appendix \ref{var-contractions}.   As for the case of $\langle \rho_4^2 \rangle_K$, things are a bit different as we have 
various angular indices $\{a, b\} = \{ \theta, \phi\}$ which should be contracted. As shown in Appendix \ref{var-contractions}, one can check that $\langle \rho_4^2 \rangle_K = 2 \langle \rho_4 \rangle_K^2$. 

Combining all results, and using the relation (\ref{rho-i-relation0}) among $\langle \rho_i \rangle_K$ we obtain
\ba
\delta \rho_K^2 &=& 2 \sum_{i=1}^3 \langle \rho_i \rangle_K^2 + \langle \rho_4^2 \rangle_K
\nonumber\\
&=& 2 \big( 4+ \frac{1}{4} +  \frac{1}{4} \big)  \langle \rho_v \rangle_K^2 +  \langle \rho_v \rangle_K^2 = 10 \langle \rho_v \rangle_K^2 \, .
\ea
Consequently, the density contrast is obtained to be
\ba
\label{contrast}
\frac{\delta \rho_K}{\langle \rho_v \rangle_K} = \pm \sqrt{10} \, .
\ea
This result is in exact agreement with the result for the density contrast in flat background obtained in \cite{Firouzjahi:2022xxb}, c.f. Eq. (\ref{density-contrast}). 
We discuss the physical implications of the above results in section \ref{implications}. 

Before closing this section we comment that we have also  calculated 
vacuum zero point energy for an observer  far from black hole, $r \gg 2 GM$.
In this limit we can neglect the gravitational effects of the black hole so the spacetime is nearly flat. Correspondingly, one expects that the vacuum zero point energy matches to its flat value. We have verified this conclusion explicitly showing that for an observer far from black hole the vacuum zero point energy is indeed 
given by Eq. (\ref{cc-formula1}).

\section{Zero point energy in dS-Schwarzschild background}

 In the previous analysis we have calculated the vacuum  zero point energy 
 in Schwarzschild black hole background  and have demonstrated that the results for $\langle \rho_v \rangle $ matches with its value in Minkowski background.  However, one can raise the valid question that in the presence of the dS vacuum energy density, the resultant background is a 
 dS-Schwarzschild spacetime. Therefore, one has to solve for the mode function in the corresponding  dS-Schwarzschild background and check if 
 Eq. (\ref{cc-formula1}) still holds. This is technically a non-trivial question, as one first has to solve for the mode function and then see if the energy density associated to these mode function yields to the starting dS energy density. This looks like moving in a loop where the final result should justify the starting assumption imposed and vice versa. 
 
 The metric for the   dS-Schwarzschild background in the static patch is given by 
 \ba
\label{dS-Sc}
ds^2 = -\Big(1- \frac{2 G M}{r} - H^2 r^2\Big) dt^2  + 
\frac{dr^2}{\Big(1- \frac{2 G M}{r} - H^2 r^2 \Big)} + r^2 d \Omega^2 \, ,
\ea
in which, as before, $M$ is the mass of the black hole and $H$ is the Hubble expansion rate of the corresponding dS background. If our analysis are consistent, then we should actually have $H = \Hm$, i.e. the corresponding dS background is entirely constructed from the vacuum zero point energy with the specific value of 
$\langle \rho_v \rangle$  given in Eq. (\ref{cc-formula1}).

The metric (\ref{dS-Sc}) has two horizons, the black hole horizon at $r_S= 2 G M$
and the cosmological horizon at $r_{dS}= H^{-1}$. To have a physical solution in which  the black hole singularity is protected behind its horizon, we require $r_S \leq r_{dS}$ so $H r_S \leq1$. The special case where the two horizons coincide with $r_S H=1$ represents the Nariai solution \cite{Nariai} in which  the spacetime becomes regular. As the two horizons coincide they have the same Hawking temperature and the system is in a thermal equilibrium:  the black hole receives the same amount of radiation from the dS space as it radiates. However, the  Nariai solution is unstable quantum mechanically \cite{Bousso:2002fq, Bousso:1997wi, Bousso:1998bn, Nojiri:1998ue, Nojiri:1998ph, Nojiri:1999pm, Nojiri:1999br}. As the two horizons are nearly identical the dynamics of quantum evaporations and the subsequent instabilities are non-trivial resulting in the fragmentation of spacetime \cite{Bousso:2002fq}.

Needless to say, it is not easy to calculate $\langle \rho_v \rangle$ in the general background of (\ref{dS-Sc}). Here, we perform the analysis of vacuum zero point energy for the particular case of Nariai solution where the calculations can be performed analytically.  
 
 The $(r, t)$ coordinate system employed in Eq. (\ref{dS-Sc}) is singular on the degenerate horizons in the Nariai limit.  Upon performing the appropriate coordinate transformation, one can show that the Nariai spacetime has the following regular line element
 \cite{Ginsparg:1982rs, Bousso:2002fq}
 \ba
 \label{Nariai}
 ds^2 = - dT^2 + \cosh^2(H T) d x^2 + \frac{1}{H^2} d \Omega^2 \, ,
 \ea
in which $T$ is a global time coordinate $-\infty < T < \infty$ (which should not be confused with the time used in Kruskal coordinate) and $x$ is a spacelike coordinate $-\infty < x < \infty$.  Also note that the radius of the remaining $S^2$ is fixed to $H^{-1}$.   The metric (\ref{Nariai}) can be viewed as a two-dimensional dS metric which is smeared over the  $S^2$. 
  
 Defining the cosmological scale factor via $a(T)= \cosh( H T)$, we  can introduce the conformal time $\tau$ via $d \tau = dT/a(T)$. This yields   $ H \tau = 2\,  \mathrm{arc} \tan (e^{H T})$  and  $e^{H T} = \tan( H \tau/2)$ with 
 $ 0< H \tau < \pi $.  Correspondingly, the scale factor in conformal time is given by
 \ba
 a( \tau) = \frac{1}{\sin ( H \tau)} \, .
 \ea
 
Using the decomposition (\ref{Phi-Z}) with $r$ fixed at $1/H$, the action is cast into
 \ba
\label{action-Nariai}
S=\frac{1}{2H^2} \int d\tau d x \sum_\ell
\left[ \,   \big | \frac{\partial Z_{\ell q}}{\partial \tau} \big|^2
- \big | \frac{\partial Z_{\ell q}}{\partial x} \big|^2 - 
\Big({ m}^2  + \ell (\ell +1) H^2 \Big) a(\tau)^2  |Z_{\ell q}|^2
\, \right] \, .
\ea 
 Expanding the mode unction in Fourier space as $Z_{\ell q} \propto e^{i kx} \varphi_{k} ^ \ell (\tau)$
 the corresponding equation for the mode function is given by
 \ba
 \partial^2_\tau \varphi_{k} ^\ell  + \left[ k^2 + \Big( { m}^2 + \ell (\ell +1) H^2 \Big) a^2 
 \right]  \varphi_{k}^ \ell  = 0 \, ,
 \ea 
in which $k \equiv | \bfk|$ while noting that $\bfk$ is one-dimensional as we Fourier expand only along one spatial direction, the $x$ direction.

 The solution of the above equation is given in terms of the hypergeometric function which is not particularly illuminating. In order to obtain a better insight into the mode function, we consider the change of coordinate $z \equiv -\cos (H \tau)$ in which 
 $-1 < z < 1$. Correspondingly, the equation of the mode function is cast into
 \ba
 \label{mode-z}
 (1- z^2) \frac{d^2  \varphi_{k}^ \ell }{ dz^2}  - z  \frac{d \varphi_{k}^\ell}{ dz}  
 + \Big[  \kappa^2 + \frac{\Delta^2}{1- z^2}  \Big] \varphi_{k}^ \ell  = 0\,  ,
 \ea
in which the dimensionless parameters $\kappa$ and $\Delta$ are defined via
\ba
\label{Delta-def}
\kappa \equiv \frac{k}{H} \, , \quad \quad
\Delta^2 \equiv \frac{{ m}^2}{H^2} + \ell (\ell + 1) \, .
\ea 
It is instructive to note that since we expect Eq. (\ref{cc-formula1}) to hold then we have $H \ll { m}$ so $\kappa, \Delta  \gg 1$.

The solution of differential equation (\ref{mode-z}) is given in terms of the associated Legendre functions $P_\nu^\mu$ and $Q_\nu^\mu$, or equivalently 
$P_\nu^\mu$ and $P_\nu^{-\mu}$ as follows
\ba
\label{sol-Legendre}
\varphi_{k}^\ell (z) = (1- z^2)^{\frac{1}{4}} \Big[ c_1 P_\nu^\mu(z) + c_2 P_\nu^{-\mu}(z) \Big] \, ,
\ea 
 in which $c_1$ and $c_2$ are two constants of normalization and the indices 
 $\nu$ and $\mu$ are defined via
 \ba
 \nu \equiv \kappa- \frac{1}{2} \, , \quad \quad \mu \equiv \frac{i}{2} ( 4 \Delta^2 - 1) ^{\frac{1}{2}} \, .
 \ea
 In particular note that $\nu \gg1$ while $\mu$ is pure imaginary with $| \mu | \gg1$.  
 
 To obtain the normalization coefficients $c_1$ and $c_2$, we have to impose the quantization conditions. We expand the quantum field $\Phi$ in terms of the one-dimensional creation and annihilation operators as follows
 \ba
\label{Z-mode-N}
\Phi = H \int \frac{d \bfk}{\sqrt{2 \pi}} \sum_{\ell q}
\frac{1}{\sqrt{2\omega(k)}}  
\Big[ e^{i k x} a_\bfk^{\ell q} \varphi_{k}^\ell (\tau) + (-1)^q e^{-i k x} {a_\bfk^{\ell\,  -q}}^\dagger \varphi_{k}^\ell (\tau)^* 
 \Big] \ylm \, ,
\ea 
in which $[ a_\bfk^{\ell q}, {a^{\ell' q'}_\bfq}^\dagger]= \delta_{\ell \ell'} \delta_{q q'} \delta (\bfk -\bfq)  $ while the other commutators all vanish.  
 
Imposing the equal time commutation relations
\ba
\label{equal-t-com}
[ \Phi( \tau, \bfx) , \partial_\tau \Phi(\tau, \bfx') ] =  i a (\tau) \delta^3 (\bfx- \bfx') \, ,
\ea 
and noting that $dz = H \sqrt{1- z^2} d \tau$, and $a= (1- z^2)^{-\frac{1}{2}}$
we obtain the following Wronskian condition on the mode function, 
\ba
\label{Wronskian}
(1- z^2) \Big( \varphi_k^\ell \, \partial_z {\varphi_k^\ell}^* - 
{\varphi_k^\ell}^* \, \partial_z {\varphi_k^\ell}  \Big) = \frac{i}{H} \, .
\ea
Plugging the mode function solution  Eq. (\ref{sol-Legendre}) into the above quantum normalization condition, we obtain\footnote{We use the Maple software to simplify the analysis concerning various relations between the associated Legendre functions, see also \cite{Abramovitz}.}
\ba
\label{c1-c2}
| c_2|^2 - | c_1|^2 =  \frac{\pi  (1- z^2)^{-\frac{1}{2}}}{2 H \sinh(\pi | \mu|)  }  \, .
\ea 
The above relation between $c_1$ and $c_2$ suggests that we can set $c_1=0$.
This is somewhat similar to the Bunch-Davies initial condition in conventional inflationary perturbations in a $1+3$ dimensional dS background in which the Bunch-Davies initial condition represents the lowest energy state.   
Of course, we can consider the general case 
where $c_1$ is not zero, corresponding to a non Bunch-Davies initial condition. However, to simplify the analysis (as dealing with the associated Legendre functions with two-indices is already challenging) we simply set $c_1=0$ in the rest of the analysis. 

Plugging the value of $c_2$ from the normalization condition (\ref{c1-c2}), the quantum mode function is given by
\ba
\label{mode-Nariai}
\Phi = \sqrt{\frac{\pi H}{2}} \int \frac{d \bfk}{\sqrt{2 \pi}} \sum_{\ell q}
\frac{1 }{\sqrt{ \sinh(\pi | \mu|) }}  
\Big[ e^{i k x} a_\bfk^{\ell q}  P_\nu^{-\mu}(z) + (-1)^q e^{-i k x} {a_\bfk^{\ell\,  -q}}^\dagger P_\nu^{\mu}(z) 
 \Big] \ylm \, .
 \ea
 
Now equipped with the above mode function, we can calculate $\langle \rho_v \rangle_N$ in the Nariai background explicitly. The subscript $N$ indicates that 
we perform the quantum expectation in the Nariai vacuum subject to the ``Bunch-Davies" type vacuum in which $c_1=0$ while $c_2$ is given by 
Eq. (\ref{mode-Nariai}). 

The vacuum energy density is given by
 \ba
 \rho_v = \frac{m^2}{2} \Phi^2 + \frac{1}{2 a(\tau)^2} \Big[  (\partial_\tau \Phi)^2 + 
  (\partial_x \Phi)^2  \Big] + \frac{H^2}{2} \gamma^{ab} \partial_a\Phi \partial_b \Phi .
 \ea
Denoting the above four contributions, as before, by $\rho_i, i=1,... 4$ we have
\ba
\label{rho1-a}
\langle \rho_1 \rangle_N &=& \frac{m^2}{2} \langle \Phi^2 \rangle 
= \frac{m^2\pi H}{4} (1- z^2)^{\frac{1}{2}} \sum_{\ell=0} \frac{2 \ell +1}{4 \pi}
\int \frac{d \bfk}{2 \pi} \frac{|P_\nu^\mu (z)|^2}{\sinh(\pi | \mu|) } , \\
\label{rho2-a}
\langle \rho_2 \rangle_N &=& \frac{1}{2 a^2} \langle  (\partial_\tau \Phi)^2  \rangle 
= \frac{\pi H^3}{4} (1- z^2)^{\frac{1}{2}} \sum_{\ell=0} \frac{2 \ell +1}{4 \pi}
\int \frac{d \bfk}{2 \pi} \frac{| {(1- z^2) P_\nu^\mu}' (z) -\frac{z P_\nu^\mu }{2 }   |^2}{\sinh(\pi | \mu|) }, \\
\label{rho3-a}
\langle \rho_3 \rangle_N &=&  \frac{1}{2 a^2} \langle  (\partial_x \Phi)^2  \rangle 
=  \frac{\pi H}{4} (1- z^2)^{\frac{3}{2}} \sum_{\ell=0} \frac{2 \ell +1}{4 \pi}
\int \frac{d \bfk}{2 \pi} k^2 \frac{|P_\nu^\mu (z)|^2}{\sinh(\pi | \mu|) }, \\
\label{rho4-a}
\langle \rho_4 \rangle_N &= &\frac{H^2}{2 }\gamma^{a b} \langle \partial_a \Phi \partial_b \Phi  \rangle 
=\frac{\pi H^3}{4} (1- z^2)^{\frac{1}{2}} \sum_{\ell=0} \frac{2 \ell +1}{4 \pi} \ell (\ell +1) 
\int \frac{d \bfk}{2 \pi}  \frac{|P_\nu^\mu (z)|^2}{\sinh(\pi | \mu|) } \, .
\ea 

Compared to analysis in Section (\ref{near-horizon}) for the near horizon region, we have the additional technical difficulties of dealing with the associated Legendre function. Fortunately, in the limit of interest where $\nu, |\mu| \gg1$, one can simplify the Legendre functions, see Appendix \ref{Legendre} for further details.  
More specifically,  
\ba
\label{Legendre-app1}
\frac{|P_\nu^\mu (z)|^2}{\sinh(\pi | \mu|) }  \simeq 
 \frac{H(1- z^2)^{\frac{1}{2}} }{\pi}  \Big[  ( 1-z^2) k^2 + \bfm^2 + \ell(\ell+1) H^2
\Big]^{-\frac{1}{2}} \, ,
\ea 
and
\ba
\label{Legendre-app2}
\frac{| {(1- z^2) P_\nu^\mu}' (z) -\frac{z P_\nu^\mu }{2 }   |^2}{\sinh(\pi | \mu|) }  \simeq 
 \frac{(1- z^2)^{\frac{1}{2}} }{H \pi}  \Big[  ( 1-z^2) k^2 + \bfm^2 + \ell(\ell+1) H^2
\Big]^{\frac{1}{2}} \, .
\ea 

Plugging  expression (\ref{Legendre-app1}) into Eq. (\ref{rho1-a}), we obtain
 \ba
 \label{rho1-b}
\langle \rho_1 \rangle_N  = \frac{2 \bfm^2 H^2}{16 \pi} 
\sum_{\ell=0} ( 2 \ell +1) ( 1- z^2)^{\frac{1}{2}} \int_0^\infty \frac{d k}{2 \pi} \Big[ ( 1-z^2) k^2 + \bfm^2 + \ell(\ell+1) H^2
\Big]^{-\frac{1}{2}} \, .
 \ea
Now, upon rescaling the wave number via $(1- z^2)^{\frac{1}{2}} k \rightarrow k$, we find that the 
expression for $\langle \rho_1 \rangle$ above matches exactly with  the corresponding value of $\langle \rho_1 \rangle$ in Eq. (\ref{rho1K}) for the near horizon region. Similarly, 
using (\ref{Legendre-app1}) in Eqs. (\ref{rho3-a}) and (\ref{rho4-a}) and rescaling 
$(1- z^2) k \rightarrow k$ we find that
the expressions  for $\langle \rho_3 \rangle$ and $\langle \rho_4 \rangle$ matches exactly with the corresponding results for $\langle \rho_3 \rangle$ and $\langle \rho_4 \rangle$ in Eqs. (\ref{rho3-1}) and (\ref{rho4-1}) respectively. Finally, employing Eq. (\ref{Legendre-app2}) into Eq. (\ref{rho2-a}) we find that the result for $\langle \rho_2 \rangle$  matches exactly with  the value of $\langle \rho_2 \rangle$ in Eq. (\ref{rho2-1}) for the near horizon region. 

Combining all contributions, we obtain 
\ba
\label{rho-val} 
\langle \rho_v \rangle_N= \sum_i^4 \langle \rho_i \rangle_N = \frac{\bfm^4}{64 \pi^2} 
\ln \Big( \frac{\bfm^2}{\mu^2} \Big) \, .
\ea
This is in exact agreement with the result in the flat  
and  Schwarzschild backgrounds. In addition, the following relations 
among $\langle \rho_i \rangle_N$, similar to Eq. (\ref{rho-i-relation0}) for the near horizon regime, hold 
\ba
\label{rho-i-relation}
\langle \rho_1 \rangle_N = 4 \langle \rho_2 \rangle_N =-4 \langle \rho_3 \rangle_N =
-2 \langle \rho_4 \rangle_N = 2 \langle \rho \rangle_N \, .
\ea
The exact agreements between $\langle \rho_i \rangle$ and $\langle \rho \rangle$
in the Nariai background with the corresponding values in the near horizon regime of the Schwarzschild background 
is reassuring. This  may also indicate the universality of the final 
results despite the technical disparities in the intermediate stages of the analysis.

Having calculated the vacuum energy density, we can now justify the approximations  $\nu, |\mu| \gg 1$ used in the above analysis. With 
$\langle \rho_v \rangle$  obtained in Eq. (\ref{rho-val}) we see that
$H= \Hm$. In other words, the dS horizon of the Nariai metric is just the dS horizon associated with the vacuum zero point energy. Furthermore, with
$\langle \rho_v \rangle \sim m^4$ we have $\Hm \sim m^2/M_P$ so
$m/\Hm \sim M_P/m \gg 1$. This indeed indicates that $\nu, |\mu| \gg 1$
so the approximations employed in  Eqs. (\ref{Legendre-app1}) and (\ref{Legendre-app2}) are very well justified.

As in near horizon regime, we can calculate the density contrast $\frac{\delta \rho_N}{\langle \rho_v \rangle_N}$ in the Nariai background in which 
$\delta \rho_N^2 \equiv \langle \rho^2 \rangle_N - \langle \rho \rangle_N^2$.   
Since the results for
$\langle \rho \rangle_N$ and $\langle \rho_i \rangle_N$ are the same as in 
 the near horizon region of the Schwarzschild background, and repeating the analysis as in subsection \ref{variance-near}, we obtain
\ba
\frac{\delta \rho_N}{\langle \rho_v \rangle_N} = \pm \sqrt{10} \, .
\ea

\section{Physical Implications}
 
 \label{implications}

Here we briefly investigate the physical implications of the results obtained in previous sections.


Associated to the vacuum energy density (\ref{cc-formula1}) there is a 
horizon radius  $\Hm^{-1}$ in which $\Hm$ is given by \cite{Firouzjahi:2022xxb}
\ba
\Hm = \Big(\frac{\langle \rho_v \rangle}{ 3 M_P^2} \Big)^{1/2} \sim \frac{\bfm^2}{M_P}\, .
\ea
The horizon radius $\Hm^{-1}$ defines a new scale which competes with the horizon scale of the black hole.  A somewhat similar situation happens in cosmological background where $\Hm^{-1}$ is compared to the FLRW horizon $H_F^{-1}$ (note that we denote the Hubble radius at the current cosmic epoch by 
$H_0^{-1}$ while the  Hubble radius at an arbitrary time in cosmic history is denoted
by $H_F^{-1}$).  Here we review some relevant results in \cite{Firouzjahi:2022xxb} in the cosmological background which can be extended to our black hole background as well. 

As argued in \cite{Firouzjahi:2022xxb} if the quantum field is light 
and $\Hm^{-1} \gg H_F^{-1}$, then the observable universe is within a single patch of the zero point energy and the effects of the vacuum energy density in cosmic expansion dynamics is minimal. However, as the universe expands then $H_F^{-1}$
increases until a time $t_m$ when the two horizon radii become comparable, 
$H_F^{-1} (t_m) \sim \Hm^{-1}$. At this time the effects of the vacuum zero point energy is not negligible. Indeed this is the time when $\delta \rho_v/\rho_T \sim 1$ 
in which $\rho_T $ is the total energy density, $\rho_T = \rho_F + \langle \rho_v \rangle$, see Eq. (\ref{fraction}). In cosmic expansion history, this happens when 
$\bfm \sim T$, in which $T$ is the background photon temperature. 
As time proceeds further and $H_F^{-1}$ increase far beyond $\Hm^{-1}$ then the density contrast becomes at the order unity and the spacetime develops strong inhomogeneities.  As strong non-linearities emerge we may not be sure of the subsequent dynamics. However, it is likely that as these overdense regions  enter the FLRW horizon they collapse to form  black holes. These black holes may play the seeds of dark matter. Alternatively, there may be more complications as 
we have both signs in density contrasts in Eq. (\ref{density-contrast}). In the regions which happen to have the plus sign in Eq. (\ref{density-contrast}) the positive energy density increases further (i.e. they become more dS-type) while regions which happen to have the negative sign in Eq. (\ref{density-contrast}) become AdS-type (note that $\sqrt{10} >1$) so they collapse to black holes. The competition between these two effects are non-trivial but the net effect may be 
described by an effective fluid with the equation of state $w_m$ with 
$-1 \leq w_m \leq 1$. The sign and the amplitude of $\omega_m$ is not clear. 
If it is close to $-1$ then we are dealing with a dark energy source while for $w_m$ close to unity we deal with a stiff fluid. 

Now let us go back to our case  of black hole background. 
 If the horizon associated to the zero point energy is very large compared to black hole event horizon, $\Hm^{-1} \gg r_S$,  then one expects that the effects of the zero point energy on black hole to be minimal. This corresponds to the case when 
$M m^2 \ll M_P^3$. This can happen when either the quantum field or the black hole are light. On the other hand, in the situation where the black hole horizon becomes comparable to the dS horizon,  $\Hm^{-1} \sim r_S$, corresponding to the case where  $M m^2 \sim M_P^3$, then one expects that the zero point energy to affect the black hole dynamics significantly.  

In the presence of the positive zero point energy  which plays the role of a cosmological constant, the spacetime is described by the dS-Schwarzschild 
metric Eq. (\ref{dS-Sc}) but with the understanding that now $H = \Hm$. 
For a fixed $r_S$, as $\Hm^{-1}$ decreases, the spacetime approaches the Nariai metric \cite{Nariai} where the two horizons coincide.  If $\Hm^{-1}$ decreases below the Nariai  limiting value, then the black hole singularity becomes naked which is not acceptable physically. 

The above discussions suggest that there is an upper bound on the mass of the
black hole immersed in a dS space created from the  zero point energy of a quantum field of mass $\bfm$. This limiting value of the black hole mass $M_N(m)$ 
in the Nariai limit  is set by the condition $r_S = \Hm^{-1}$, yielding
\ba
\label{mass-limit} 
M_N(m) = \frac{4 \pi \sqrt3 M_P^3}{\sqrt{\langle \rho_v \rangle}}  = \frac{32 \pi^2 \sqrt3}{\Big[\ln 
\big( \frac{\bfm^2}{\mu^2} \big) \Big]^{\frac{1}{2}}} \, \frac{M_P^3}{\bfm^2} \, .
\ea  

There is an interesting interpretation for the above limiting mass formula. Motivated by discussions in \cite{Firouzjahi:2022xxb} (as described at the start of this subsection) suppose we encounter the situation where the condition  $\frac{\delta \rho}{\rho} \sim 1$ has been met so the dS space filled with the zero point energy starts to fragment. Now suppose  the Hubble radius of size $\Hm^{-1}$ collapse to form a black hole. Denoting the mass of this black hole by $ {M}_B(m)$ we find
 \ba
\label{MB(m)}
 M_B(m) &=& \frac{4 \pi}{3}   \langle  \rho_v\rangle \Hm^{-3}  
= \frac{32 \pi^2 \sqrt3 }{\Big[ \ln( \frac{\bfm^2}{\mu^2}) \Big]^{\frac{1}{2}}} \frac{M_P^3}{\bfm^2} \, .
\ea
Interestingly, we find that $ M_B(m) =  M_N(m)$. In other words, the limiting mass in the Nariai metric is nothing but the mass enclosed in a sphere of radius 
$\Hm^{-1}$ with the energy density $ \langle  \rho_v\rangle$ furnished by the vacuum zero point energy. 
This result supports the proposal suggested in \cite{Firouzjahi:2022xxb} that the region inside the dS horizon of zero point energy may collapse into a black hole 
but now with the additional understanding that the resulting black hole is the Nariai solution with the maximum allowed mass.


Assuming there is no large hierarchy between the renormalization scale $\mu$ and $m$ in Eq. (\ref{mass-limit}), we obtain the following simple estimation for the black hole upper mass
\ba
\label{mass-limit2} 
M_B(m) \sim 10^2 \frac{M_P^3}{\bfm^2} \, .
\ea
 Interestingly, the above formula is basically the same as the 
Chandrasekhar limiting mass formula\footnote{We thank Misao Sasaki for bringing this similarity to our attention.} for astrophysical compact objects 
though here we have obtained it in a very different way.

\section{Summary and Discussions}

In this paper we have studied the vacuum zero point energy in a black hole background. We are interested in the limit where the Compton length of the fundamental field is much smaller than the black hole horizon radius, corresponding to $M \gg M_P^2/m$.  We have  demonstrated the validity of 
 Eq. (\ref{cc-formula1}) in a black hole background, both for the near horizon and 
 for the far observer. In addition, we have demonstrated the validity of Eq. (\ref{cc-formula1}) in the Nariai background as a specific limit of the dS-Schwarzschild background.  Our result is inline with conclusion in \cite{Martin:2012bt} where the validity 
of Eq. (\ref{cc-formula1}) was demonstrated for a general curved background.

Although the validity of  Eq. (\ref{cc-formula1})  in a curved background 
might have been expected based on reasonings from equivalence principle  but the actual demonstration of its validity in a black hole background is non-trivial. This is because the black hole setup has a horizon which separates the interior singularity from the outer space. As such, the notion of vacuum is a non-trivial concept in which different observers are equipped with different vacua. However,  based on equivalence principle, only the vacuum defined by a locally free falling observer  is expected to be Lorentz invariant. 
As we have shown, for an observer near the black hole horizon, the vacuum defined in the Kruskal coordinate has the advantage that it  is non-singular on the horizon (and everywhere). We have demonstrated that the vacuum energy density as measured by this local inertial observer indeed matches with Eq. (\ref{cc-formula1}). In addition, we have calculated the vacuum energy density for an observer far from black hole in which the effects of the black hole mass is negligible. For this observer the spacetime is asymptotically Minkowskian so there is no surprise that  Eq. (\ref{cc-formula1}) should hold true for this case  as well. 

In the presence of vacuum zero point energy, the spacetime will be deformed from a pure Schwarzschild background and one actually deals with a dS-Schwarzschild setup. Therefore, in principle, we have to calculate the vacuum zero point energy in this new background. Needless to say, the calculations for this more complicated setup can not be done  analytically. Therefore, we have restricted ourselves to the Nariai setup where the black hole  and the cosmological backgrounds 
have equal horizon radius. 
The Nariai background has interesting properties which were studied in the past, for example it represents a thermal equilibrium where the black hole and the dS  horizons share the same temperature. Although the analysis of the mode function were non-trivial but we were able to demonstrate that Eq. (\ref{cc-formula1}) holds true in the Nariai background as well.

Motivated by the analysis of  \cite{Firouzjahi:2022xxb} we have also calculated the density contrast associated  to the zero point energy. We have checked that indeed $\delta \rho/ \langle \rho_v \rangle =\pm \sqrt{10}$ both in the black hole  and in the Nariai setups.  
As argued in \cite{Firouzjahi:2022xxb} this reflects the instability of the dS spacetime created from the vacuum quantum fluctuations. One can imagine that parts of the spacetime where $\delta \rho/ \langle \rho_v \rangle$ takes the negative sign are AdS-type which consequently may collapse to black holes. Now the crucial question is what the mass of the resulting black holes is? 
We have shown that the mass of the resulting black hole is given by the upper mass limit in the Nariai metric which is equal to the mass enclosed in a patch of  dS horizon constructed from the vacuum zero point energy.  This  supports  the conjecture made in  \cite{Firouzjahi:2022xxb} that the space filled with the vacuum zero point energy is unstable to fragmentation and may form 
black holes. 

To simplify the analysis, we have performed the calculations for the case of a real scalar field. Physically we expect the result to be extended to other types of  fundamental fields such as the fermions or vector bosons. The main difference would be that the  vacuum energy density and its sign will depend on the spin and the polarization degrees of freedom of the corresponding field as outlined in \cite{Martin:2012bt}.
In addition, we expect that only massive fields to contribute to the vacuum zero point energy while massless fields such as graviton or photon make no contribution.

\vspace{1cm}
   
 {\bf Acknowledgments:}  We would like to thank Paul Anderson, 
 Mohammad Ali Gorji, Raihaneh Moti, Haidar Sheikhahmadi and Alireza Talebian  for insightful comments and  discussions. 
  
 \vspace{0.5cm}
  
\appendix 
\section{Regularizations of infinite sum}
\label{Sum-reg}

As part of obtaining the regularized vacuum zero point energy for near horizon regime, we need to evaluate
the infinite sum of the following form
\ba
\label{Sp}
{\bf S}_{(p)}(x) \equiv \sum_{\ell =0}^{\infty} ( 2\ell +1) \Big( x+ \ell (\ell+1) \Big)^{-p} \, ,
\ea
where in the notation of Eq. (\ref{rho1-1}),  $p= \frac{2-d}{2}$ and $x= \kappa^2$.
The above sum converges for $p>1$. But in our case, we analytically continue the result for negative value of $p$ to obtain the finite physical result.

\begin{figure}[t]
	\centering
	\includegraphics[ width=0.6\linewidth]{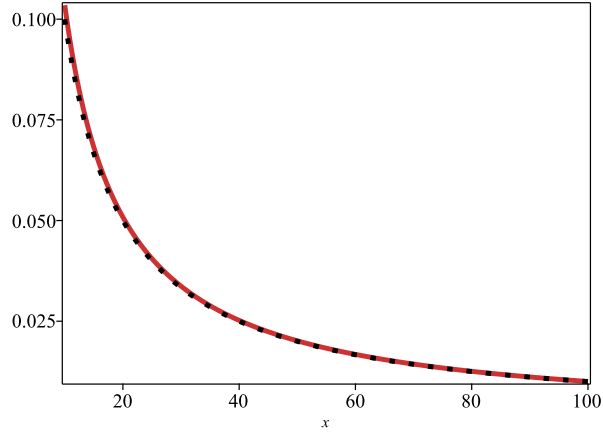}
	\vspace{-6 cm}
	\caption{ The plots of the function  ${\bf S}_{(p)}(x)$ and its approximate form 
given in Eq. (\ref{Sp-approximation}) for $p=2$. The approximation Eq. (\ref{Sp-approximation}) is valid with the accuracy better than one percent for
$x \gg p$.  }
\label{Sp-plot}
\end{figure}

As we discussed in subsection \ref{near-horizon} we are interest in the physical limit where the Compton radius of the quantum field is much smaller than the black hole horizon, corresponding to $\kappa \equiv m r_s \gg1$. 
We have checked numerically that in this limit of interest where  $x \gg p$, ${\bf S}_{(p)}(x) $ is very well approximated by
\ba
\label{Sp-approximation}
{\bf  S}_{(p)}(x)  \simeq \frac{x^{1-p}}{p-1} \, .
\ea
The accuracy of the approximation is typically better than one percent. 
In Fig. \ref{Sp-plot} the plots of the function ${\bf S}_{(p)}(x)$ and its approximation 
in Eq. (\ref{Sp-approximation}) are presented, confirming the accuracy of the 
approximate result Eq. (\ref{Sp-approximation}). 

For the case $\langle \rho_1 \rangle $ given in Eq. (\ref{rho1-1}) we have 
$p= \frac{2-d}{2}$ yielding  Eq. (\ref{rho1}).  
For the cases of $\langle \rho_2 \rangle $ and $\langle \rho_3 \rangle $ given in Eqs. (\ref{rho2-1}) and  (\ref{rho3-1}) we have
$p= -\frac{d}{2}$, yielding to  Eqs. (\ref{rho2}) and  (\ref{rho3}) respectively. 

On the other hand, in calculating $\langle \rho_4 \rangle $ we have encountered the following sum
\ba
\hat{\bf S}_{(q)}(x) \equiv \sum_{\ell=0}^{\infty} ( 2 \ell + 1) \ell (\ell+1)  \Big( x + \ell (\ell+1) \Big)^{-q}   \, .
\ea
We can relate $\hat{\bf S}_{(q)}$ into our known sum ${\bf S_{(p)}}(x)$ by the following trick
\ba
(1-q) \hat{\bf S}_{(q)}(x) &=& \frac{d}{d \lambda} \Big[ \sum_{\ell=0}^{\infty} (2 \ell+1) 
\Big( x+ \lambda \ell (\ell+1)  \Big)^{1-q} \Big] \Big|_{\lambda=1} \nonumber\\
&=& \frac{d}{d \lambda} \Big[  \lambda^{1-q}\sum_{\ell=0}^{\infty} (2 \ell+1) 
\Big( \frac{x}{\lambda}+ \ell (\ell+1)  \Big)^{1-q} \Big] \Big|_{\lambda=1} \, .
\ea
The sum inside the last big bracket above is  ${\bf S}_{(q-1)}(\frac{x}{\lambda})$. Now using our approximation given in Eq. (\ref{Sp-approximation})  we obtain
\ba
(1-q) \hat{\bf S}_{(q)}(x)=  \frac{d}{d \lambda} \Big(  \lambda^{1-q}{\bf S}_{(q-1)}(\frac{x}{\lambda})\Big) \Big|_{\lambda=1}  \simeq \frac{d}{d \lambda} \Big(  \lambda^{-1} \frac{x^{2-q}}{q-2} \Big) \Big|_{\lambda=1} = \frac{x^{2-q}}{2-q} \, .
\ea
For the sum involving $\langle \rho_4 \rangle $ in Eq. (\ref{rho4-1}) we have
$q=\frac{2-d}{2}$ and $x= \kappa^2$, yielding Eq. (\ref{hatS}).

As discussed after Eq. (\ref{rho-i-relation0}) in order to calculate the finite terms in $\langle \Phi^2 \rangle$ in the context of vacuum polarization, we have to calculate the next subleading corrections in powers of $\kappa^{-1}$ in $S_1$. Examining the numerical results, the corrections in $S_1$ up to $m^{-4}$ 
have the following form
\ba
{\bf S}_{(p)}(x) \simeq \frac{x^{1-p}}{p-1} \Big[ 1+ \frac{(p-1)}{3 x} 
+ \frac{p (p-1)}{15 x^2} \Big]  \, .
\ea
Using the above expression in calculating $S_1$ in Eq. (\ref{S1-def}) and discarding the terms containing $m^2, m^0$ and $\ln(m)$, the leading finite term
in  $\langle \Phi^2 \rangle$ at $r= r_S$ is obtained to be 
\ba
\langle \Phi^2 \rangle= \frac{1}{240 \pi^2} \frac{1}{m^2 r_S^4}\, ,
\ea
in agreement with \cite{Frolov:1982fr, Anderson:1990jh}. 

\section{Higher order contractions in variance analysis}
\label{var-contractions}

In this Appendix we present the analysis of higher order  contractions
relevant for the variance analysis in section \ref{variance-near}. 
We restrict ourselves to the near horizon region with the mode function 
\ba
\label{Z-mode-near-b}
\Phi = \frac{1}{r_S} \int \frac{d \bfk}{\sqrt{2 \pi}} \sum_{\ell q}
\frac{1}{\sqrt{2\omega(k)}}  
\Big[ e^{i k\cdot x} a_\bfk^{\ell q} + (-1)^q e^{-i k\cdot x} {a_\bfk^{\ell\,  -q}}^\dagger
 \Big] \ylm \, .
\ea 

As mentioned in the main text, since the field $\Phi$ is Gaussian one expects that the Wick theorem to apply and in particular 
\ba
\label{Wick1}
\langle  \rho_i^2 \rangle_K = 3  \langle  \rho_i \rangle_K^2   ,\quad \quad
i=1, 2, 3,
\ea
in which 
\ba
\rho_1 \equiv \frac{1}{2} m^2 \Phi^2 , \quad
\rho_2 \equiv  \frac{1}{2} (\partial_\tau \Phi)^2 , \quad
\rho_3 \equiv \frac{1}{2} (\partial_x \Phi)^2 , \quad
\rho_4 \equiv  \frac{1}{2 r_S^2} \gamma^{a b} \partial_a \Phi \partial_b \Phi \, .
\ea
Note that since $\rho_4$ carries the contractions of angular indices $a, b$ then 
it turns out that $\langle  \rho_4^2 \rangle_K = 2  \langle  \rho_4 \rangle_K^2$.
Here we outline the derivations of these results more specifically. 

To demonstrate (\ref{Wick1}) we perform the analysis for the case of  $\langle  \rho_2^2 \rangle_K$ as an example. The analysis for the cases $i=1, 3$ are very similar. We also remove the subscript $K$ for convenience. 

Using the mode function given in Eq. (\ref{Z-mode-near-b})
we have
\ba
\label{contractions}
\langle  \rho_2^2 \rangle = \frac{-1}{4r_S^4} \sum_{\ell_j, q_j}  \prod_{j=1}^4
&&Y_{\ell_j q_j} (\theta, \phi) 
\int \frac{d \bfk_j}{\sqrt{2 \pi}}  \sqrt{\omega_j}   \times
\\
&& \langle \,  a_{\bfk_1}^{\ell_1 q_1} \big( a_{\bfk_2}^{\ell_2 q_2} - (-1)^{q_2} {a_{\bfk_2}^{\ell_2 \, -q_2}}^{\dagger}   \big) \big( a_{\bfk_3}^{\ell_3 q_3} - (-1)^{q_3} {a_{\bfk_3}^{\ell_3 \, -q_3}}^{\dagger}   \big) (-1)^{q_4}  {a_{\bfk_4}^{\ell_4 \, -q_4}}^{\dagger} \,   \rangle .  \nonumber
\ea
Note that $-\ell_j \leq q_j \leq \ell_j$ for $j=1...4$ in the above sum. 

There are two types of contributions $A$ and $B$ in $\langle  \rho_2^2 \rangle$ as follows:
\ba
A &\equiv& \frac{1}{4r_S^4} \sum_{\ell_j, q_j}  \prod_{j=1}^4
(-1)^{q_3} (-1)^{q_4}
Y_{\ell_j q_j}   \int \frac{d \bfk_j}{\sqrt{2 \pi}} \sqrt{ \omega_j }
\langle \,  a_{\bfk_1}^{\ell_1 q_1}  a_{\bfk_2}^{\ell_2 q_2}
 {a_{\bfk_3}^{\ell_3\, - q_3}}^{\dagger}  {a_{\bfk_4}^{\ell_4\, - q_4}}^{\dagger}   \rangle\\
 &=& \frac{2}{4r_S^4} \sum_{\ell_j, q_j}  \prod_{j=1}^4
(-1)^{q_3} (-1)^{q_4}
Y_{\ell_j q_j}   \int \frac{d \bfk_j}{\sqrt{2 \pi}}  \sqrt{\omega_j }
\delta_{\ell_1 \ell_3} \delta_{\ell_2 \ell_4} \delta_{q_1\,  -q_3} \delta_{q_2\,  -q_4}
\delta(\bfk_1 -\bfk_3) \delta(\bfk_2 -\bfk_4) , \nonumber
\ea 
where the factor 2 comes from the other permutation replacing $3\leftrightarrow 4$ above. 

Performing the integral over the $\delta$ function we obtain
\ba
A= \frac{2}{4r_S^4} \Big(\sum_{\ell q} |Y_{\ell q}|^2   \int \frac{d \bfk}{\sqrt{2 \pi}}  \omega \Big)^2 = 2 \langle \rho_2 \rangle^2 \, .
\ea

The other contribution from the contractions in Eq. (\ref{contractions}) is given by
\ba
B &\equiv& \frac{1}{4r_S^4} \sum_{\ell_j, q_j}  \prod_{j=1}^4
(-1)^{q_2} (-1)^{q_4}
Y_{\ell_j q_j}   \int \frac{d \bfk_j}{\sqrt{2 \pi}}  \sqrt{\omega_j }
\langle \,  a_{\bfk_1}^{\ell_1 q_1}  {a_{\bfk_2}^{\ell_2 \, -q_2}}^\dagger
 {a_{\bfk_3}^{\ell_3\ q_3}}  {a_{\bfk_4}^{\ell_4\, - q_4}}^{\dagger}   \rangle\\
 &=& \frac{1}{4r_S^4} \sum_{\ell_j, q_j}  \prod_{j=1}^4
(-1)^{q_2} (-1)^{q_4}
Y_{\ell_j q_j}   \int \frac{d \bfk_j}{\sqrt{2 \pi}} \sqrt{ \omega_j }
\delta_{\ell_1 \ell_2} \delta_{\ell_3 \ell_4} \delta_{q_1\,  -q_2} \delta_{q_3\,  -q_4}
\delta(\bfk_1 -\bfk_2) \delta(\bfk_3 -\bfk_4) , \nonumber
\ea 
which yields
\ba
B= \frac{1}{4r_S^4} \Big(\sum_{\ell q} |Y_{\ell q}|^2   \int \frac{d \bfk}{\sqrt{2 \pi}}  \omega \Big)^2 =  \langle \rho_2 \rangle^2 \, .
\ea
Adding the two combinations  $A+B$ we obtain $\langle  \rho_2^2 \rangle = 3  \langle  \rho_2 \rangle^2$ as promised. 

The case of $\langle  \rho_4^2 \rangle$ is somewhat similar but there are new effects from the contractions of the angular indices $\{a, b\} = \{ \theta, \phi \}$. To see this explicitly, 
let us start with 
\ba
\langle  \rho_4^2 \rangle = \frac{1}{4 r_S^4}\gamma^{a b} \gamma^{cd} 
\langle  \nabla_a  \Phi\nabla_b  \Phi   \nabla_c  \Phi\nabla_d  \Phi
\rangle \, .
\ea
As the field is Gaussian, we can perform the contractions and there are two different contributions, $C$ and $D$ as follows:
\ba
C \equiv   \frac{1}{4 r_S^4}\gamma^{a b} \gamma^{cd} 
\langle  \nabla_a  \Phi\nabla_b  \Phi   \rangle \langle  \nabla_c  \Phi\nabla_d  \Phi
\rangle \, ,
\ea
and
\ba
D\equiv   \frac{2}{4 r_S^4}\gamma^{a b} \gamma^{cd} 
\langle  \nabla_a  \Phi\nabla_c  \Phi   \rangle \langle  \nabla_b  \Phi\nabla_d  \Phi
\rangle \, ,
\ea
where the factor 2 comes from two identical permutations $b \leftrightarrow c$. 

Our job is now to determine the allowed form of $\langle  \nabla_a  \Phi\nabla_b  \Phi   \rangle$. Since this expression is symmetric under the change of its two indices, it should be proportional to the metric of the two-sphere $\gamma_{ab}$ as
$\gamma_{ab}$ is the only covariant symmetric tensor available on this background. Therefore,
\ba
\langle  \nabla_a  \Phi\nabla_b  \Phi   \rangle \equiv  \lambda \gamma_{ab} \, .
\ea
To find the proportionality factor $\lambda$ we contract it with the inverse metric 
$\gamma^{ab}$ obtaining
\ba
\lambda = \frac{1}{2} \gamma^{a b} \langle  \nabla_a  \Phi\nabla_b  \Phi   \rangle 
= r_S^2 \langle \rho_4 \rangle. 
\ea
Now plugging this value for the expressions in $C$ and $D$ we obtain 
$D= C = \langle  \rho_4 \rangle^2$ yielding in total 
\ba
\langle  \rho_4^2 \rangle = 2 \langle  \rho_4 \rangle^2 \, .
\ea


\section{Approximation of the Associated Legendre Functions}
\label{Legendre}

In this Appendix, we present the approximate relation for the associated Legendre functions appearing in $\langle  \rho_i^2 \rangle$ in the Nariai background. 

We work in the limit where $\nu, |\mu| \gg 1$. In this limit, one can check that 
\ba
\label{Legendre-app1b}
\frac{|P_\nu^\mu (z)|^2}{\sinh(\pi | \mu|) }  \simeq 
\frac{(1- z^2)^{\frac{1}{2}} }{\pi} \Big[ \Delta^2 + ( 1-z^2) (\nu+ \frac{1}{2})  
\Big]^{-\frac{1}{2}} 
\ea 
in which $\mu = \frac{i}{2} ( 4 \Delta^2 - 1) ^{\frac{1}{2}}$. 

\begin{figure}[t]
	\centering
	\includegraphics[ width=0.48\linewidth]{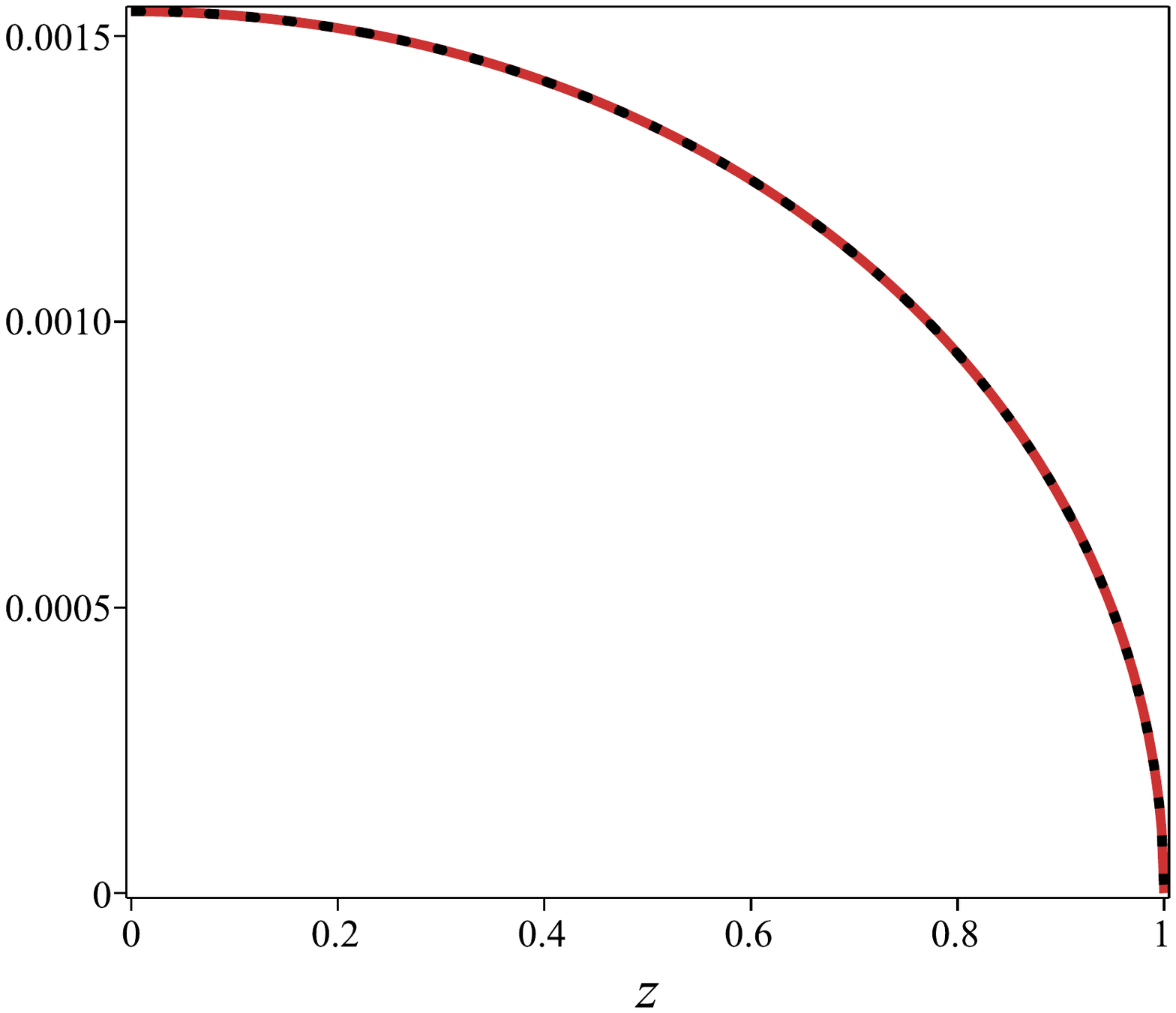}
	\includegraphics[ width=0.48\linewidth]{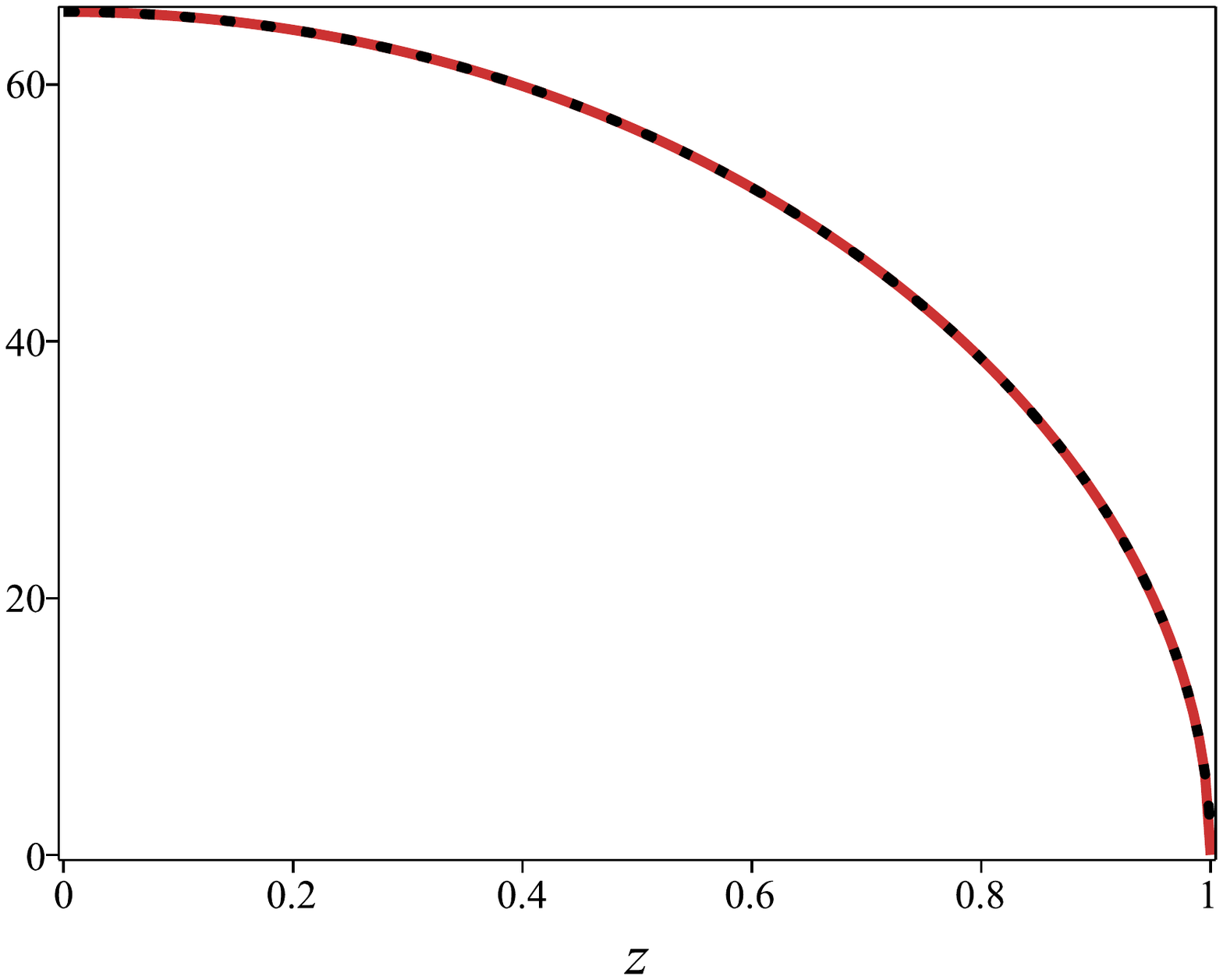}
	\vspace{-3.5 cm}
	\caption{ The plots of the approximation relations  Eq. (\ref{Legendre-app1b}) (Left) and  Eq. (\ref{Legendre-app2b}) (Right).  In both plots we have set $\nu=50$ and $\Delta=200$. The agreement between the exact functions
	in left hand sides of Eqs. (\ref{Legendre-app1b}) and (\ref{Legendre-app2b}) 
	and their corresponding approximations are extremely good where $\nu, \Delta \gg 1$.  
 }
\label{Legendre12-plot}
\end{figure}

We have checked that the approximation (\ref{Legendre-app1b}) is very accurate, as can be seen in the left panel of Fig. \ref{Legendre12-plot}. 
Using the definitions of $\nu$ and $\Delta $ from Eq. (\ref{Delta-def}) we finally have
\ba
\label{Legendre-app1c}
\frac{|P_\nu^\mu (z)|^2}{\sinh(\pi | \mu|) }  \simeq 
 \frac{H(1- z^2)^{\frac{1}{2}} }{\pi}  \Big[  ( 1-z^2) k^2 + \bfm^2 + \ell(\ell+1) H^2 \Big]^{-\frac{1}{2}} \, .
\ea 

Similarly, we have checked that the following approximation holds to very high accuracy 
\ba
\label{Legendre-app2b}
\frac{| {(1- z^2) P_\nu^\mu}' (z) -\frac{z P_\nu^\mu }{2 }   |^2}{\sinh(\pi | \mu|) }  &\simeq& 
\frac{(1- z^2)^{\frac{1}{2}} }{H\pi} \Big[ \Delta^2 + ( 1-z^2) (\nu+ \frac{1}{2})  
\Big]^{\frac{1}{2}} \, , \\
 &=&\frac{(1- z^2)^{\frac{1}{2}} }{H \pi}  \Big[  ( 1-z^2) k^2 + \bfm^2 + \ell(\ell+1) H^2
\Big]^{\frac{1}{2}} \, ,
\ea 
as can bee seen in the right panel of Fig. \ref{Legendre12-plot}.

\vspace{0.5cm}

{}


\begin{thebibliography}{}

\bibitem{Weinberg:1988cp}
S.~Weinberg,
``The Cosmological Constant Problem,''
Rev. Mod. Phys. \textbf{61}, 1-23 (1989).


\bibitem{Sahni:1999gb}
V.~Sahni and A.~A.~Starobinsky,
Int. J. Mod. Phys. D \textbf{9}, 373-444 (2000).

\bibitem{Peebles:2002gy}
P.~J.~E.~Peebles and B.~Ratra,
Rev. Mod. Phys. \textbf{75}, 559-606 (2003).


\bibitem{Copeland:2006wr}
E.~J.~Copeland, M.~Sami and S.~Tsujikawa,
Int. J. Mod. Phys. D \textbf{15}, 1753-1936 (2006).


\bibitem{Martin:2012bt}
J.~Martin,
Comptes Rendus Physique \textbf{13}, 566-665 (2012),
[arXiv:1205.3365 [astro-ph.CO]].

\bibitem{Weinberg:2008zzc}
S.~Weinberg,
``Cosmology,'' Oxford University Press, 2008. 


\bibitem{Planck:2018vyg}
N.~Aghanim \textit{et al.} [Planck],
Astron. Astrophys. \textbf{641}, A6 (2020)
[erratum: Astron. Astrophys. \textbf{652}, C4 (2021)]. 


\bibitem{Firouzjahi:2022xxb}
H.~Firouzjahi,
[arXiv:2201.02016 [gr-qc]].






\bibitem{Akhmedov:2002ts}
E.~K.~Akhmedov,
[arXiv:hep-th/0204048 [hep-th]].

\bibitem{Koksma:2011cq}
J.~F.~Koksma and T.~Prokopec,
[arXiv:1105.6296 [gr-qc]].

\bibitem{Ossola:2003ku}
G.~Ossola and A.~Sirlin,
Eur. Phys. J. C \textbf{31}, 165-175 (2003). 

\bibitem{Visser:2016mtr}
M.~Visser,
Particles \textbf{1}, no.1, 138-154 (2018). 

\bibitem{Weinberg:1995mt}
S.~Weinberg,
``The Quantum theory of fields. Vol. 1: Foundations.''






\bibitem{Abbott:2016blz} 
  B.~P.~Abbott {\it et al.} [LIGO Scientific and Virgo Collaborations],
  Phys.\ Rev.\ Lett.\  {\bf 116}, no. 6, 061102 (2016).

\bibitem{Abbott:2016nmj} 
  B.~P.~Abbott {\it et al.} [LIGO Scientific and Virgo Collaborations],
  Phys.\ Rev.\ Lett.\  {\bf 116}, no. 24, 241103 (2016).
  


\bibitem{Wang:2017oiy}
Q.~Wang, Z.~Zhu and W.~G.~Unruh,
Phys. Rev. D \textbf{95}, no.10, 103504 (2017). 

\bibitem{Cree:2018mcx}
S.~S.~Cree, T.~M.~Davis, T.~C.~Ralph, Q.~Wang, Z.~Zhu and W.~G.~Unruh,
Phys. Rev. D \textbf{98}, no.6, 063506 (2018). 

\bibitem{Wang:2019mbh}
Q.~Wang and W.~G.~Unruh,
Phys. Rev. D \textbf{102}, no.2, 023537 (2020). 

\bibitem{Wang:2019mee}
Q.~Wang,
Phys. Rev. Lett. \textbf{125}, no.5, 051301 (2020). 

\bibitem{Birrell:1982ix}
N.~D.~Birrell and P.~C.~W.~Davies,
``Quantum Fields in Curved Space.''


\bibitem{Parker:2009uva}
L.~E.~Parker and D.~Toms,
``Quantum Field Theory in Curved Spacetime: Quantized Field and Gravity.''

\bibitem{Moreno-Pulido:2020anb}
C.~Moreno-Pulido and J.~Sola,
Eur. Phys. J. C \textbf{80}, no.8, 692 (2020).

\bibitem{Moreno-Pulido:2021jmn}
C.~Moreno-Pulido and J.~S.~Peracaula,
[arXiv:2110.08070 [gr-qc]].


\bibitem{Moreno-Pulido:2022phq}
C.~Moreno-Pulido and J.~S.~Peracaula,
[arXiv:2201.05827 [gr-qc]].






\bibitem{Candelas:1980zt}
P.~Candelas,
Phys. Rev. D \textbf{21}, 2185-2202 (1980). 

\bibitem{Howard:1984qp}
K.~W.~Howard and P.~Candelas,
Phys. Rev. Lett. \textbf{53}, 403-406 (1984).

\bibitem{Frolov:1982fr}
V.~P.~Frolov and A.~I.~Zelnikov,
Phys. Lett. B \textbf{115}, 372-374 (1982).

\bibitem{Frolov:1983ig}
V.~p.~Frolov and A.~i.~Zelnikov,
Phys. Lett. B \textbf{123}, 197-199 (1983).


\bibitem{Anderson:1990jh}
P.~R.~Anderson,
Phys. Rev. D \textbf{41}, 1152-1162 (1990).

\bibitem{Anderson:1994hg}
P.~R.~Anderson, W.~A.~Hiscock and D.~A.~Samuel,
Phys. Rev. D \textbf{51}, 4337-4358 (1995), 

\bibitem{DeWitt:1975ys}
B.~S.~DeWitt,
Phys. Rept. \textbf{19}, 295-357 (1975). 

\bibitem{Fulling:1989nb}
S.~A.~Fulling,
``Aspects of Quantum Field Theory in Curved Space-time.''

\bibitem{Christensen:1976vb}
S.~M.~Christensen,
Phys. Rev. D \textbf{14}, 2490-2501 (1976). 

\bibitem{Christensen:1978yd}
S.~M.~Christensen,
Phys. Rev. D \textbf{17}, 946-963 (1978). 




\bibitem{Watson}
G. N. Watson,
A Treatise on the Theory of Bessel Functions, Second Edition, 1944,
pp 152. 



\bibitem{Hawking:1974sw} 
  S.~W.~Hawking,
  Commun.\ Math.\ Phys.\  {\bf 43}, 199 (1975)
  Erratum: [Commun.\ Math.\ Phys.\  {\bf 46}, 206 (1976)].


\bibitem{Weinberg:1995mt}
S.~Weinberg,
``The Quantum theory of fields. Vol. 1: Foundations,''
Cambridge University Press, 2005.




\bibitem{Nariai} 
H. Nariai, 
``On a new cosmological solution of Einstein's field equations of gravitation," 
Sci. Rep. Tohoku Univ. Ser. I 35, 62 (1951).

\bibitem{Bousso:2002fq}
R.~Bousso,
[arXiv:hep-th/0205177 [hep-th]].

\bibitem{Bousso:1997wi}
R.~Bousso and S.~W.~Hawking,
Phys. Rev. D \textbf{57}, 2436-2442 (1998).

\bibitem{Bousso:1998bn}
R.~Bousso,
Phys. Rev. D \textbf{58}, 083511 (1998).


\bibitem{Nojiri:1998ue}
S.~Nojiri and S.~D.~Odintsov,
Int. J. Mod. Phys. A \textbf{14} (1999), 1293-1304. 

\bibitem{Nojiri:1998ph}
S.~Nojiri and S.~D.~Odintsov,
Phys. Rev. D \textbf{59} (1999), 044026.

\bibitem{Nojiri:1999pm}
S.~Nojiri and S.~D.~Odintsov,
Int. J. Mod. Phys. A \textbf{15} (2000), 989-1010.

\bibitem{Nojiri:1999br}
S.~Nojiri and S.~D.~Odintsov,
Phys. Lett. B \textbf{463} (1999), 57-62.


\bibitem{Ginsparg:1982rs}
P.~H.~Ginsparg and M.~J.~Perry,
Nucl. Phys. B \textbf{222}, 245-268 (1983). 


\bibitem{Abramovitz}
M. Abramowitz and I. Stegun,
``Handbook of Mathematical Functions with Formulas, Graphs, and Mathematical Tables," 1972. 








\end{thebibliography}
\end{document}